\documentclass[a4paper,twocolumn,accepted=2025-02-25]{quantumarticle}
\pdfoutput=1
\usepackage[utf8]{inputenc}
\usepackage[english]{babel}
\usepackage[T1]{fontenc}

\usepackage{mdframed}
\usepackage{booktabs}
\usepackage{longtable}

\usepackage[section]{placeins}
\usepackage{amsfonts}
\usepackage{amssymb}
\usepackage{physics}
\usepackage{hyperref}
\usepackage[mathlines]{lineno}
\usepackage{mathtools}
\usepackage{nccmath}
\usepackage{float}
\usepackage{appendix}
\usepackage{dcolumn}
\usepackage{bm}
\usepackage[numbers,sort&compress]{natbib}
\usepackage{orcidlink}

\newtheorem{definition}{Definition}

\renewcommand{\j}{\mathrm{j}}
\newcommand{\vi}{\mathrm{i}}
\renewcommand{\k}{\mathrm{k}}

\newcommand{\m}{\mathrm{m}}
\newcommand{\n}{\mathrm{n}}
\newcommand{\C}{\mathcal{C}}
\newcommand{\CI}{\mathcal{C}_\textrm{\normalfont{I}}}
\newcommand{\CD}{\mathcal{C}_\textrm{\normalfont{D}}}
\newcommand{\CR}{\mathcal{C}_\textrm{\normalfont{R}}}
\newcommand{\DeltaI}{\Delta_\textrm{\normalfont{I}}}
\newcommand{\DeltaD}{\Delta_\textrm{\normalfont{D}}}
\newcommand{\DeltaR}{\Delta_\textrm{\normalfont{R}}}
\newcommand{\CIprox}{\mathcal{C}_{{\textrm{\normalfont{I}}}}}
\newcommand{\CDprox}{\mathcal{C}_{{\textrm{\normalfont{D}}}}}

\begin{document}
\title{Wavefunction branching: when you can't tell pure states from mixed states}
\author{Jordan K. Taylor}
\email[]{jordantensor@gmail.com}
\affiliation{School of Mathematics and Physics, University of Queensland, Brisbane, Queensland 4072, Australia}
\orcid{0000-0002-5799-0557}
\author{Ian P. McCulloch}
\affiliation{School of Mathematics and Physics, University of Queensland, Brisbane, Queensland 4072, Australia}
\affiliation{Department of Physics, National Tsing Hua University, Hsinchu 30013, Taiwan}
\affiliation{Frontier Center for Theory and Computation, National Tsing Hua University, Hsinchu 30013, Taiwan}
\orcid{0000-0002-8983-6327}
\maketitle
\onecolumn{
\begin{abstract}
\vspace{-0.3cm}
\noindent
We propose a definition of wavefunction ``branchings'': quantum superpositions which can't be feasibly distinguished from the corresponding mixed state, even under time evolution. Our definition is largely independent of interpretations, requiring only that it takes many more local gates to swap branches than to distinguish them. We give several examples of states admitting such branch decompositions. 
Under our definition, we argue that attempts to get relative-phase information between branches will fail without frequent active error correction, that branches are effectively the opposite of good error-correcting codes,  that branches effectively only grow further apart in time under natural evolution, that branches tend to absorb spatial entanglement, that branching is stronger in the presence of conserved quantities, and that branching implies effective irreversibility. 
Identifying these branch decompositions in many-body quantum states could shed light on the emergence of classicality, provide a metric for experimental tests at the quantum/ classical boundary, and allow for longer numerical time evolution simulations. We see this work as a generalization of the basic ideas of environmentally-induced decoherence to situations with no clear system/ environment split.
\end{abstract}}
\vspace{1.5cm}
\twocolumn\
\vspace{-2.0cm}
\tableofcontents

\section{\label{sec:intro}Introduction}
\begin{figure}
    \centering
    \includegraphics[width=\linewidth]{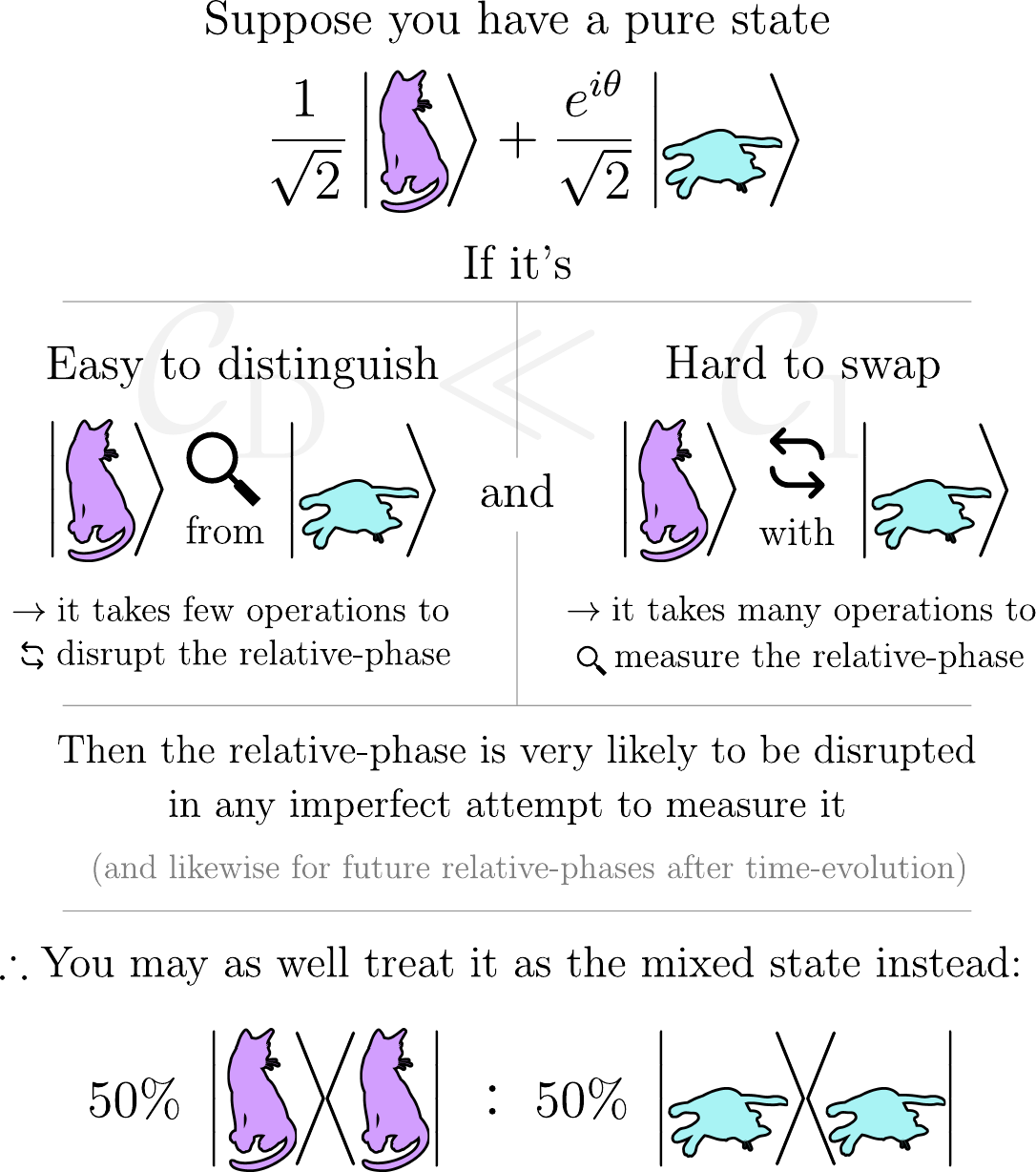}
    \caption{A summary of the core idea that we propose and build upon in this paper.}
    \label{fig:Branches_theory_summary}
\end{figure}
Consider the growth of entanglement over time in many-body quantum systems.
For a wide range of initial states and local Hamiltonians, entanglement entropy between a local subsystem and its complement grows approximately linearly with time until it reaches a value proportional to the subsystem size~\cite{calabrese2005evolution, lauchli2008spreading, kim2013ballistic, kaufman2016quantum,  alba2017entanglement, nahum2017quantum, zhou2020entanglement, de2022linear, bertini2022growth}. When taken to the thermodynamic limit, this implies a generic unrelenting growth of entanglement with time.

This is a problem for numerical simulation techniques such as tensor networks, which have algorithmic costs scaling exponentially with bipartite entanglement entropy. Numerical time evolution simulations often run across an ``entanglement barrier'', where the cost of simulating these systems grows exponentially with time simulated~\cite{schuch2008hardness}. Considered from another angle however, the necessity of modeling this unrelenting entanglement growth in large quantum systems is puzzling; macroscopic phenomena tend to be well approximated by classical physics. May we approximate the full quantum state at some time by, say, a classical probability distribution of less entangled states, such that all feasible experiments won't be able to tell the difference? A scheme like this may allow much longer time evolution simulations by using sampling methods to overcome the entanglement barrier, as suggested by Riedel~\cite{riedel2017classical}.

Decoherence is a phenomenon by which quantum-mechanical systems can become well described by classical probability distributions~\cite{anastopoulos2002FAQdecoherence}. Typically, a preferred factorisation of Hilbert space into a system and an environment is assumed. However in the case of generic pure-state closed-system or translationally-invariant time evolution, it is often unclear where to draw the system/environment distinction, if anywhere.

Instead, we take as our starting point arbitrary pure quantum states of isolated closed systems, without assuming any preferred decomposition into system and environment. We want a principled way of looking for decompositions of such a state $|\Psi\rangle$ into components $|\psi_\vi \rangle$ (sometimes called ``branches'')
\begin{equation}\label{eq:branch_decomp}
|\Psi\rangle = \sum_\vi  c_\vi | \psi_\vi \rangle
\end{equation}
such that differences between the pure density matrix $\rho_{} \coloneqq |\Psi\rangle \langle\Psi| = \sum_{\vi \j } c_\vi  c^*_\j  |\psi_\vi \rangle\langle\psi_\j |$ and a classical probability distribution of branches $\rho_{\text{diag}} = \sum_{\vi } |c_\vi |^2 |\psi_\vi \rangle\langle\psi_\vi |$ won't be revealed under further evolution or feasible quantum operations. In these cases the quantum state $\ket{\Psi}$ can be replaced with a classical probability distribution over the branches.

Better understanding the conditions by which pure states can be treated as classical probability distributions like this should also shed light on the nature of the quantum to classical transition. This problem of adequately defining such ``branch decompositions'' is related to other problems of identifying the emergence of apparent classicality in quantum theory, such as the ``set-selection problem'' in the context of consistent histories~\cite{dowker1996consistent}, the ``preferred factorisation problem'' or ``quantum mereology'' in the context of decoherence~\cite{carroll2021quantum}, and the ``quantum reality problem'' more generally~\cite{kent2014solution}.  

Some inspirational recent work in this area has included expanding the formalism of decoherence to closed systems~\cite{fortin2014decoherenceClosed}, identifying preferred system and environment distinctions~\cite{carroll2021quantum}, analyzing the circuit complexity required to detect macroscopic superpositions~\cite{aaronson2020hardness}, defining good branches by identifying redundant local records of information~\cite{riedel2016objective, riedel2017classical}, and defining good branches by minimizing the ``net complexity'' of the branch decomposition~\cite{weingarten2022macroscopic}. Our approach is also constructed in terms of quantum complexity, but the complexity of interference and distinguishability experiments between branches, focused on the conditions under which $\rho$ can be replaced with $\rho_\text{diag}$. We summarize this approach in figure \ref{fig:Branches_theory_summary}.

\section{\label{sec:complexity}A complexity based definition of branches}
We take a pragmatic view of branches: One may talk about a wavefunction in terms of branches whenever this would give a more convenient description of any potentially feasible experiments.

What do we mean by ``feasible?'' Quantum complexity may be a useful tool here, even though it does not necessarily correspond to feasibility. The circuit complexity  $\C(U)$ of some unitary operation $U$ is the minimum number of primitive operations (usually one and two-qubit gates) required to perform it.\footnote{
 For an intuitive introduction to circuit complexity as a measure of distance between states, see ref~\cite{susskind2018three_lectures}.
 For concreteness in this paper we allow arbitrary two-qubit gates. We do not restrict the gates to be spatially local (between adjacent degrees of freedom). However the details of all of these choices do not significantly affect what counts as a good branch in the examples we consider (except for the GHZ state, where we explicitly note otherwise).
 Likewise, there are many other definitions and generalizations of complexity, such as Nielsen's continuous geometric generalization~\cite{nielsen2006quantum}. We expect the determination of which systems have good branch decompositions to generalize across different complexity measures so long as they capture total circuit size rather than just depth.  Future work should use any definition of complexity which best captures the allowed primitive operations in the relevant systems of interest. }

High-complexity operations are those which involve operations on many separate degrees of freedom, take a long time to implement at a fixed energy scale, or both. Operations with high complexity are usually infeasible to perform precisely without some form of quantum error correction, because small perturbations are generally magnified exponentially in complexity~\cite{brown2017quantum_complexity_curvature, dalzell2021random_circuits_local_noise_into_global_brandao}. These perturbations may come from an inability to implement the desired quantum circuit to infinite precision, for example.

\begin{figure}
    \centering
    \includegraphics{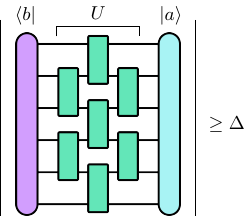}
    \caption{The relative-state complexity $\CR(\ket{a},\ket{b},\Delta)$ between two states $|{{a}}\rangle$ and $|{{b}}\rangle$ is the minimum number of gates $\C(U)$ in any circuit $U$ satisfying this inequality. The circuit will approximately map from $|{{a}}\rangle$ to $|{{b}}\rangle$, up to a phase.} 
    \label{fig:CR}
\end{figure}

As well as operations, complexity can also be measured between states. Most common is the relative-state complexity. For our purposes, this is the minimum number of 1 or 2-qubit gates required to approximately map one state to another up to a phase: 
\begin{definition}\label{def:CR}The relative state complexity $\CR\left(\ket{a}, \ket{b}, \Delta \right)$ is the minimum number of gates in any circuit $U$ satisfying $\left|\langle b | {U} |a \rangle\right| \geq \Delta$,
\end{definition}
where $\Delta$ is a number between $0$ and $1$ capturing the accuracy required of the circuit: when $\Delta=1$, the circuit must perfectly map $\ket{a}$ to $e^{i\theta}\ket{b}$. This definition is illustrated in figure \ref{fig:CR}. 

Recent conjectures have proposed a ``second law of quantum complexity'' for the relative-state complexity $\CR\left(|\psi(0)\rangle,\ |\psi(t)\rangle,\ \Delta \right)$, stating that the complexity of mapping between an initial state $|\psi(0)\rangle$ and a later state $|\psi(t)\rangle$ generally only increases with time, in analogy with the classical entropy of an exponentially larger system under the second law of thermodynamics~\cite{brown2017quantum_complexity_curvature,brown2018second, halpern2021resource_theory_complexity}. This was a key inspiration for our work, as we think that complexity is a relevant measure by which branches may become well defined and separated over time, in analogy with the growth of entropy and thermodynamic irreversibility in classical systems. 

However we give our definition of good branch decompositions in terms of two alternate measures of complexity between states: the complexity $\CI$ of detecting interference between branches, and the complexity $\CD$ of distinguishing between branches. We will say that proposed branch decompositions are valid when $\CI-\CD$ is large; when it is much harder to detect interference between them than to distinguish them. 
\subsection{Distinguishability complexity}\label{sec:distinguishability-complexity}
The complexity of distinguishing between two states $|a\rangle$ and $|b\rangle$ has been studied by Aaronson, Atia, and Susskind, who found it to be essentially the same as the complexity of swapping $\ket{\psi_+} = \frac{|a\rangle + |b\rangle}{\sqrt{2}}$ with $\ket{\psi_-} = \frac{|a\rangle - |b\rangle}{\sqrt{2}}$~\cite{aaronson2020hardness}. In a nutshell, distinguishing two states requires essentially the same number of gates as swapping their relative-phases in a superposition. 

\begin{figure}
    \centering
    \includegraphics{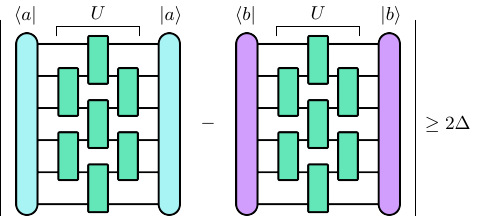}
    \caption{The distinguishability complexity proxy $\CDprox(\ket{a},\ket{b},\Delta)$ between two states $|{{a}}\rangle$ is the minimum number of gates in any circuit $U$ satisfying this inequality. The circuit $U$ will be able to $\Delta$-approximately swap $\frac{|{{a}}\rangle+e^{i\theta}|{{b}}\rangle}{\sqrt{2}}$ with $\frac{|{{a}}\rangle-e^{i\theta}|{{b}}\rangle}{\sqrt{2}}$ at any phase $\theta$. The complexity $\CDprox$ of this circuit is within a constant factor plus $O(1)$ gates of the true complexity $\CD$ of distinguishing $|{{a}}\rangle$ from $|{{b}}\rangle$ with a difference in probabilities of at least $\Delta$~\cite{aaronson2020hardness}. }
    \label{fig:CD_simplified}
\end{figure}

The extent to which a circuit $U$ swaps two states $\ket{\psi_+}$ and $\ket{\psi_-}$ can be quantified by $\left|\langle \psi_+| U | \psi_- \rangle + \langle \psi_-| U | \psi_+ \rangle \right|/2$: this is only equal to $1$ if $U$ maps $\ket{\psi_+}$ to $\ket{\psi_-}$ and $\ket{\psi_-}$ to $\ket{\psi_+}$.  For any unitary $U$, a test can be constructed discriminating between $|a\rangle$ and $|b\rangle$ with a probability difference of $\left|\langle \psi_+| U | \psi_- \rangle + \langle \psi_-| U | \psi_+ \rangle \right|/2$ = $\left|\langle a| U | a \rangle - \langle b| U | b \rangle \right|/2$, at a circuit complexity of $O[\C(U)]$~\cite{aaronson2020hardness}, motivating a proxy measure of distinguishability complexity:
\begin{definition}\label{def:CD_proxy}\label{def:CD}The distinguishability complexity proxy $\CDprox(\ket{a}, \ket{b},\Delta)$ is the minimum number of gates in any circuit $U$ satisfying $\left|\langle a| U | a \rangle - \langle b| U | b \rangle \right| \geq 2\Delta$,
\end{definition}
as illustrated in figure \ref{fig:CD_simplified}.
This distinguishability complexity proxy is useful because it is easy to work with, while always remaining within a constant factor plus $O(1)$ gates of the true distinguishability complexity~\cite{aaronson2020hardness}. A perfect distinguishability-proxy circuit could, for example, map $\ket{a}$ to $\ket{a}$ and $\ket{b}$ to $-\ket{b}$.

\subsection{Interference complexity}\label{sec:interference-complexity}
An interference experiment is anything which can provide information on the relative phase between states. If we are given a superposition $\frac{|a\rangle + e^{i\phi}|b\rangle}{\sqrt{2}}$ with some unknown relative phase $\phi$, being able to distinguish between $\frac{|a\rangle + e^{i\theta}|b\rangle}{\sqrt{2}}$ and $\frac{|a\rangle - e^{i\theta}|b\rangle}{\sqrt{2}}$ at any relative-phase $\theta$ would provide information constraining the true relative-phase $\phi$. 
As such, we define the interference complexity as the minimum complexity of distinguishing $\frac{|a\rangle + e^{i\theta}|b\rangle}{\sqrt{2}}$ from $\frac{|a\rangle - e^{i\theta}|b\rangle}{\sqrt{2}}$ at any relative-phase $\theta$:
\begin{equation*}
    \CI(|a \rangle,|b \rangle,\Delta) \coloneqq \min_\theta\left[\CD\left(\frac{|a\rangle + e^{i\theta}|b\rangle}{\sqrt{2}},\frac{|a\rangle - e^{i\theta}|b\rangle}{\sqrt{2}},\Delta\right)\right].
\end{equation*}
A perfect interference circuit would map $\ket{a}$ and $\ket{b}$ into states which both overlap as much as possible with the same product state. 

\begin{figure}
    \centering
    \includegraphics{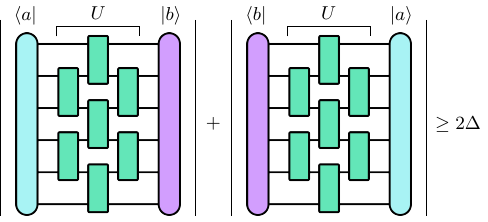}
    \caption{The interference complexity proxy $\CIprox(\ket{a},\ket{b},\Delta)$ is the minimum number of gates in any circuit $U$ satisfying this inequality ($\Delta$-approximately swapping $\ket{a}$ and $\ket{b}$ up to a phase). It is within a constant factor plus $O(1)$ gates of the true interference complexity $\CI(\ket{a},\ket{b},\Delta)$ of discriminating between $\frac{|a\rangle + e^{i\theta}|b\rangle}{\sqrt{2}}$ and $\frac{|a\rangle - e^{i\theta}|b\rangle}{\sqrt{2}}$ at some easiest phase $\theta$~\cite{aaronson2020hardness}. }
    \label{fig:CI_simplified}
\end{figure}

We can find a simpler form for the interference complexity by again noting that the complexity of discriminating between $\frac{|a\rangle + |b\rangle}{\sqrt{2}}$ and $\frac{|a\rangle - |b\rangle}{\sqrt{2}}$ is essentially the same as the complexity of swapping $|a\rangle$ with $|b\rangle$~\cite{aaronson2020hardness}. Getting relative-phase information between two states requires essentially the same number of gates as swapping them. Specifically, for any unitary $U$ a test can be constructed discriminating between $\frac{|a\rangle +|b\rangle}{\sqrt{2}}$ and $\frac{|a\rangle -|b\rangle}{\sqrt{2}}$ with a probability difference of $\left|\langle a| U | b \rangle + \langle b| U | a \rangle \right|/2$, at a circuit complexity of $O[\C(U)]$~\cite{aaronson2020hardness}. We can use this to define an alternate measure of the interference complexity by maximizing this probability difference over any relative-phase:
\begin{equation}
\begin{split}
    &\max_\theta\left|\langle a| U \left(e^{i\theta}| b \rangle\right) +  \left(e^{-i\theta}\langle b |\right)  U | a \rangle \right| \\
    & =  \left|\langle a| U | b \rangle\right| + \left|\langle b| U | a \rangle \right|,
\end{split}    
\end{equation}
so we define the interference complexity proxy as
\begin{definition}\label{def:CI_proxy}\label{def:CI}The interference complexity proxy $\CIprox(|a \rangle,|b \rangle,\Delta)$ is the minimum number of gates in any circuit $U$ satisfying $\left|\langle a| U | b \rangle\right| + \left|\langle b| U | a \rangle \right| \geq 2\Delta$, 
\end{definition}
as illustrated in figure \ref{fig:CI_simplified}.
This interference complexity proxy is simpler to calculate and intuit, while always remaining within a constant factor plus $O(1)$ gates of the true interference complexity~\cite{aaronson2020hardness}. A perfect interference-proxy circuit would map $\ket{a}$ to $\ket{b}$ up to a phase, and $\ket{b}$ to $\ket{a}$ up to a phase. 
See appendix \ref{sec:misc_properties} for more properties of the interference and distinguishability complexities.


\subsection{Good branch decompositions}\label{sec:good_branches}
Consider decomposing any state $\ket{\Psi}$ into two orthogonal components $\ket{a}$ and $\ket{b}$:
\begin{equation}|\Psi\rangle=\sqrt{p_a}|a\rangle + e^{i\theta}\sqrt{p_b}|b\rangle,\label{eq:Psi_decomposed_two_states}
\end{equation}
where $p_a$ and $p_b$ are real numbers such that $p_a+p_b=1$. The density matrix of this pure state is
\begin{equation}
\begin{split}
    \rho(\theta)=|\Psi\rangle\langle\Psi| &=  p_a\ketbra{a}{a} +  e^{-i\theta}\sqrt{p_a p_b}\ketbra{a}{b} \\ &+  e^{i\theta}\sqrt{p_a p_b}\ketbra{b}{a} + p_b\ketbra{b}{b},
\end{split}\label{eq:rho_ab}
\end{equation}
whereas the corresponding ``branched'' diagonal density matrix is
\begin{align}
    \rho_\text{diag}= p_a\ketbra{a}{a} + p_b\ketbra{b}{b} 
         = \frac{1}{2\pi} \int_0^{2\pi} \rho(\theta) \, d\theta,\label{eq:rho_ab_diag}
\end{align}
representing a classical probability distribution over the states $\ket{a}$ and $\ket{b}$, rather than a single pure state $\ket{\Psi}$. 

If it will be infeasible to ever discriminate between $\rho(\theta, t)$ and $\rho_\text{diag}(t)$ at any future time $t$ for any relative-phase $\theta$, then the proposed branch decomposition in equation (\ref{eq:Psi_decomposed_two_states}) is a good one, and $\rho_\text{diag}$ may be used instead of $\rho(\theta)$.

Due to linearity, any difference in outcome probabilities between $\rho(\theta)$ and $\rho_\text{diag}(t)$ can written entirely in terms of the pure states $\frac{|a\rangle \pm e^{i\theta}|b\rangle}{\sqrt{2}}$: 
\begin{equation}
\begin{split}
&\left|P\left(m\,|\,U,\,\rho(\theta)\right) - P\left(m\,|\,U,\,\rho_\text{diag}\right)\right|/\sqrt{p_a p_b}\\
= &\  \left|P\left(m\,|\,U,\ \rho(\theta)-\rho_\text{diag}\ \right)\right|/\sqrt{p_a p_b}\\
= \ & \left|P\left(m\,|\,U,\frac{\ket{a} + e^{i\theta}\ket{b}}{\sqrt{2}}\right) - P\left(m\,|\,U,\frac{\ket{a} - e^{i\theta}\ket{b}}{\sqrt{2}} \right)\right|,\label{eq:interference_bias_pure}
\end{split}
\end{equation}
so a high interference complexity between $\ket{a}$ and $\ket{b}$ means that discriminating $\rho(\theta)$ from $\rho_\text{diag}$ will require a large circuit, regardless of $\theta$. If the interference complexity is high even for getting tiny probability differences, then a large circuit will be required to get any relative-phase information whatsoever. In a nutshell, $\CI\left(\ket{a}, \ket{b}, \epsilon \right)$ should be large in order for $\ket{a}$ and $\ket{b}$ to be good branches, where $0<\epsilon \ll 1$.

A large interference complexity means that a large circuit would be required to invalidate the replacement of $\rho(\theta)$ with $\rho_\text{diag}$, but just because something requires a large circuit doesn't mean that doing it is infeasible. In good error-correcting codes for example, even though the interference complexity may be large, relative-phase information may still be feasible to obtain. So high interference complexity alone is not sufficient for a definition of good branches. We want a condition where even very long time evolution will not be sufficient to reveal differences 
between $\rho(\theta)$ and $\rho_\text{diag}$.

We argue that in order for a branching to be meaningful, it should be feasible to tell one branch from another; the distinguishability complexity between $\ket{a}$ and $\ket{b}$ should be low. This idea is reminiscent of quantum Darwinism and other definitions of branches based on local records, where a key criteria for branching is the presence of distinct, redundantly stored locally accessible information in each branch ~\cite{riedel2017classical}. Low distinguishability complexity is similar in spirit, except that it relaxes the requirement that the information be redundantly stored (as we investigate in appendix \ref{sec:robustness}). Easy distinguishability is also important for excluding error-correcting codes, which don't make good branches because they can protect relative-phase information from scrambling and allow it to be measured even if $\CI$ is large. Low distinguishability complexity prevents good error-correcting codes from being called good branches, as we will see in section \ref{sec:error_correction}. 

Putting these two requirements together, we say that a proposed decomposition of a state $|\Psi\rangle$ as a sum of orthogonal components is a good branch decomposition if the interference complexity is much higher than the distinguishability complexity between each of the branches: 

\begin{definition}~\label{def:valid_branch} A decomposition $|\Psi\rangle = \sum_\vi  c_\vi | \psi_\vi \rangle$, is a good branch decomposition if  $\CI(|\psi_\vi \rangle,|\psi_\j \rangle,\epsilon)-\CD(|\psi_\vi \rangle,|\psi_\j \rangle,1-\epsilon) \gg 1$ for all $\vi\neq \j$, where $\epsilon$ is a sufficiently small constant ($\ll \frac{1}{2}$). For simplicity of analysis in this paper, we also enforce that branches are exactly orthogonal: $\braket{\psi_\vi}{\psi_\j } = \delta_{\vi \j }$. 
\end{definition}
Even if we did not enforce exact orthogonality in this definition, it would still imply that branches must be nearly orthogonal: $\langle \psi_\vi|\psi_\j\rangle\leq \epsilon$ $\forall\  \vi\neq\j$, but for the purposes of convenient analysis in this paper we enforce that all proposed branch decompositions are exactly orthogonal. We expect this condition could be relaxed back to approximate orthogonality without damaging the spirit of our results.  Additionally, definition \ref{def:valid_branch} implies that $\CI(|\psi_\vi \rangle,|\psi_\j \rangle,\DeltaI)-\CD(|\psi_\vi \rangle,|\psi_\j \rangle,\DeltaD)$ is large for any $\DeltaI$ between $\epsilon$ and $1-\epsilon$, and any $\DeltaD$ between $\epsilon$ and $1-\epsilon$, because $\CI(|\psi_\vi \rangle,|\psi_\j \rangle,\Delta)$ and $\CD(|\psi_\vi \rangle,|\psi_\j \rangle,\Delta)$ are monotonically increasing functions of $\Delta$.


We argue in sections \ref{sec:error_correction} through to \ref{sec:mixed_states} that this definition is sufficient to allow replacing wavefunctions with probability distributions over branches, so long as there is no ``adversarial'' Maxwell's-demon-like situation where active error correction is applied sufficiently frequently and accurately in a directed attempt to reduce the interference complexity and get relative-phase information between branches.

The analogy to Maxwell's demon arises because of Brown and Susskind's  conjecture linking quantum complexity to classical thermodynamics~\cite{brown2018second}. In their ``second-law of quantum complexity'', a correspondence is proposed between the circuit complexity of a system of $N$ qubits, and the entropy of a classical system with $2^N$ degrees of freedom. An agent trying to reduce $\CR(\ket{\psi(t)},\ket{0}^{\otimes N},\Delta)$ over time is then equivalent to Maxwell's demon trying to violate the second law of thermodynamics in an exponentially larger classical system of size $2^N$. 

We propose a definition of ``adversarially robust'' branches which should stand up even to (imperfect) demons:
\begin{definition}\label{def:great_branch}$|\Psi\rangle = \sum_\vi  c_\vi | \psi_\vi \rangle$ is an adversarially robust branch decomposition if  $\CI(|\psi_\vi \rangle,|\psi_\j \rangle,\epsilon) > \exp\left[{\lambda  \CD(|\psi_\vi \rangle,|\psi_\j \rangle,1-\epsilon)}\right]$ for all $\vi\neq \j$ where $\epsilon$ is sufficiently small ($\ll \frac{1}{2}$) and $\lambda\propto\ln({1/p})$ is some sufficiently large parameter quantifying the minimum feasible noise rate $p$ per round of error correction.
\end{definition}
We argue in section \ref{sec:adversarial_robustness} that getting \textbf{current} relative-phase information between such ``adversarially robust'' branches can be infeasible even with frequent active error correction playing the role of a sort of Maxwell's demon. Though, as we will see in section \ref{sec:time_evolution}, such a Maxwell's demon may still be able to get \textbf{future} relative-phase information between branches: although good branches and adversarially-robust branches both remain good branches over time, adversarial robustness is not necessarily preserved.

\section{Examples}\label{sec:examples}
The easiest way to understand what counts as a good branch decomposition (definition  \ref{def:valid_branch}) is by looking at examples.
\subsection{GHZ state}\label{sec:example:GHZ}
Consider the obvious branch decomposition of a GHZ state:
\begin{equation}\alpha|000\cdots\rangle + \beta|111\cdots\rangle \longrightarrow \left[\;\alpha|000\cdots\rangle,\; \beta|111\cdots\rangle \;\right], \end{equation}
where $|000\cdots\rangle$ and $|111\cdots\rangle$ are spatial product states.
Here the interference complexity increases with the system size because swapping $|000\cdots\rangle$ with $|111\cdots\rangle$ requires an extensive number of gates (e.g. one $Z$ gate for each qubit). So 
\begin{equation} \CI = O(N),  \end{equation}
where $N$ is the system size. On the other hand, distinguishing $|000\cdots\rangle$ from $|111\cdots\rangle$ (or equivalently, swapping $\alpha|000\cdots\rangle + \beta|111\cdots\rangle$ with $\alpha|000\cdots\rangle - \beta|111\cdots\rangle $) only requires a single operation on one qubit. Therefore
\begin{gather}
\CD = 1,\end{gather} so this may be a good branch decomposition when $N\gg1$. However the interference-proxy circuit (definition \ref{def:CI_proxy}) only requires single-qubit gates, and the true interference circuit (definition \ref{def:CI}) only requires identical CNOT Clifford gates. These gates are sometimes not counted with the same weight as general two-qubit gates, because they can fail to generate some measures of complexity. If these complexity measures are adopted, there may be no finite system size $N$ for which a simple GHZ state truly forms good branches. In this sense, although a GHZ state has the prototypical ``Schr\"odinger's cat'' entanglement pattern that one might associate with the Von-Neumann measurement model~\cite{hobson2022entanglement_measurement}, it may be a borderline case for true branching due to its lack of internal complexity.

\subsection{Product state + random state}\label{sec:example:product_random}
A much more difficult state in which to detect interference is a superposition of a product state $|000\cdots\rangle$ and a Haar-random state $|\eta\rangle$. Consider the obvious branching of this superposition:
\begin{equation}\alpha|000\cdots\rangle + \beta|\eta\rangle \longrightarrow \left[\;\alpha|000\cdots\rangle,\; \beta|\eta\rangle \;\right]. \end{equation}
Haar random states generally take exponentially many gates to prepare from (or swap with) product states such as $|000\cdots\rangle$~\cite{knill1995approximation}, so 
\begin{gather}\CI \approx O(\mathrm{exp}(N))\end{gather}
whereas $|000\cdots\rangle$ can be distinguished from a random state with high probability just by looking at a constant number of qubits, meaning that  
\begin{gather}
\CD = O(1).\end{gather}
So this can be a good branch decomposition if $\exp(N)\gg1$. Here we get good branch decompositions at exponentially smaller system sizes than with a GHZ state. 

\subsection{Different random circuits}\label{sec:example:random_circuits_depths}
In a generalization of the example in section \ref{sec:example:product_random}, let $|\gamma_i\rangle$ be a product state to which a random quantum circuit of depth $D_i$ (where $1 \ll D_i \ll \mathrm{exp}(N)$) has been applied everywhere, and consider branching a superposition of two different such states
\begin{equation}
    \alpha|\gamma_1\rangle + \beta|\gamma_2\rangle \longrightarrow \left[\; \alpha|\gamma_1\rangle, \;  \beta|\gamma_2\rangle \;\right],
\end{equation}
so that we recover section \ref{sec:example:product_random} if $D_1 \rightarrow 0$ and $D_2 \rightarrow \infty$. The interference complexity $\CI(|\gamma_1\rangle,\, |\gamma_2\rangle,\, \epsilon)$ is the number of gates required to $\epsilon$-accurately swap $|\gamma_1\rangle $ with $|\gamma_2\rangle $. If $\epsilon\ll\frac{1}{2}$, we can settle for a circuit which maps $|\gamma_1\rangle$ to $|\gamma_2\rangle $ while just mapping $|\gamma_2\rangle$ to something mostly orthogonal to $|\gamma_1\rangle $. This can usually be done by applying the inverse of the circuit which created $|\gamma_1\rangle$, then re-implementing the circuit which created $|\gamma_2\rangle$. In fact, it has been shown that there is usually no way of mapping $|\gamma_1\rangle$ to $|\gamma_2\rangle$ using significantly fewer gates than this~\cite{haferkamp2022linear}. Therefore, the interference complexities add:
\begin{gather}
\CI(|\gamma_1\rangle,\, |\gamma_2\rangle,\, \epsilon) \approx O\left((D_1 + D_2) N\right).\end{gather}
In order to distinguish $|\gamma_1\rangle$ from $|\gamma_2\rangle$, however, we only need to apply the inverse of one of the generating circuits, then check if we get back to the initial $|000\cdots\rangle$ state. Again, there is usually no alternate circuit requiring many fewer gates than this~\cite{haferkamp2022linear}. So
\begin{gather}
    \CD \approx O\left( \mathrm{min}(D_1, D_2) N\right).\end{gather}
Clearly this is likely to be a good branch decomposition when the circuits are very unbalanced: if $D_1 \gg D_2$ or $D_2 \gg D_1$ and $N$ is large. 

However even when the circuits are of equal depth, $D_1=D_2=D$, $\CI \approx O(2DN)$ is still larger than $\CD \approx O(DN)$ by a factor of two, which can satisfy the definition for good branches (definition \ref{def:valid_branch}) if $N\times D$ is very large. 

The most general condition for this example to be a good branch decomposition is:
\begin{equation}
    (D_1 + D_2 - \mathrm{min}(D_1,D_2))N \gg 1
\end{equation}
or equivalently 
\begin{equation}
    \mathrm{max}(D_1,D_2)N \gg 1
\end{equation}
Therefore superpositions of sufficiently large random circuits (supported on the same spatial region) make good branches. 

\subsection{Surface codes}\label{sec:example:toric}
\begin{figure}
    \centering
    \includegraphics{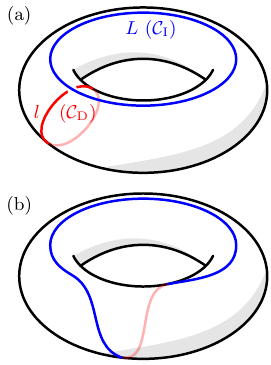}
\caption{(a) In the $L\times l$ toric code $\CI\propto L$ and $\CD \propto l$ between two logical codestates $|\bar0\rangle$ and $|\bar1\rangle$. Getting relative-phase information between $|\bar0\rangle$ and $|\bar1\rangle$ requires applying $L$ gates around the torus, whereas affecting the relative phase takes $l$ gates. (b) Logical relative-phase errors can occur if the interference circuit accidentally wraps around the wrong dimension of the torus. This becomes more likely as $L$ ($\CI$) is increased or $l$ ($\CD$) is decreased. }
    \label{fig:toric_code}
\end{figure}

Let $|\bar0\rangle$ and $|\bar1\rangle$ be the two logical code states of the $L\times l$ rectangular surface code~\cite{lee2021rectangular_surface_code_toric}, and consider the branching 
\begin{gather}
    \alpha|\bar0\rangle + \beta|\bar1\rangle \longrightarrow \left[\; \alpha|\bar0\rangle, \;  \beta|\bar1\rangle \;\right]. 
\end{gather}
In this case,
\begin{gather}    
\CI = O\left(L\right),\\
\CD = O\left(l\right),\end{gather}
so this is only a good branch decomposition if $L\gg l$.  Figure \ref{fig:toric_code} illustrates why this happens for the analogous case of the rectangular toric code. On the other hand, a superposition in the conjugate basis $\alpha|\bar+\rangle + \beta|\bar-\rangle \longrightarrow \left[\; \alpha|\bar+\rangle, \;  \beta|\bar-\rangle \;\right] $ is only a good branching if $l\gg L$. Good branches do not exist in the codespace if the surface is square rather than rectangular: if $L\approx l$. Surface codes are discussed in more detail in section \ref{sec:adversarial_robustness}. 

\subsection{Quantum parity code}\label{sec:example:parity}
The quantum parity code is a generalization of the Shor code, encoding one logical qubit in $m_1\times m_2$ physical qubits. The logical states are:
\begin{equation}|\bar{0}\rangle = \frac{1}{2^{m_2/2}} \left(|0\rangle^{\otimes m_1} + |1\rangle^{\otimes m_1} \right)^{\otimes m_2}, \end{equation}
\begin{equation}|\bar{1}\rangle = \frac{1}{2^{m_2/2}} \left(|0\rangle^{\otimes m_1} - |1\rangle^{\otimes m_1} \right)^{\otimes m_2}.\end{equation}
For this code, $ \CD\left(|\bar{0}\rangle, |\bar{1}\rangle \right) = m_1 $ and $\CI \left(|\bar{0}\rangle, |\bar{1}\rangle \right) = m_2$.
The codewords $[|\bar{0}\rangle, \, |\bar{1}\rangle]$ only make good branches if $m_2 \gg m_1$, while $[|\bar{0}\rangle + |\bar{1}\rangle,\, |\bar{0}\rangle - |\bar{1}\rangle]$ only make good branches if $m_1 \gg m_2$. Good branches do not exist in the codespace if $m_1 \approx m_2$. See more explanation of the relationship between branches and error correcting codes in section \ref{sec:error_correction}.

\subsection{Separable systems}\label{sec:example:separable}
Consider a superposition of two states $|\psi_L\rangle $ and $|\phi_L\rangle $ in some region $L$, along with a separate isolated state $|R\rangle$:
\begin{equation}(|\psi_L\rangle + |\phi_L\rangle) \otimes |R\rangle   \longrightarrow \left[ |\psi_L\rangle \otimes |R\rangle,\, |\phi_L\rangle \otimes |R\rangle \right].\end{equation}
We would expect that this branching in $L$ should be unaffected by the extraneous system $R$. Indeed, 
\begin{gather}\CI( |\psi_L\rangle \otimes |R\rangle,\, |\phi_L\rangle \otimes |R\rangle) = \CI( |\psi_L\rangle,\, |\phi_L\rangle),\\
\CD(|\psi_L\rangle \otimes |R\rangle,\, |\phi_L\rangle \otimes |R \rangle)=\CD( |\psi_L\rangle,\, |\phi_L\rangle).\end{gather}
So branchings which would occur in isolated systems still occur regardless of what is happening in other isolated systems. 

\subsection{Entangled systems}\label{sec:example:entangled}
Systems may be entangled in the original wavefunction but not in each of the branches:
\begin{equation}|\psi_L\rangle \otimes |\psi_R\rangle + |\phi_L\rangle \otimes |\phi_R\rangle   \longrightarrow \left[ |\psi_L\rangle \otimes |\psi_R\rangle,\, |\phi_L\rangle \otimes |\phi_R\rangle \right].\end{equation}
Here we find that entanglement causes $\CI$ and $\CD$ to behave differently:
\begin{align}\begin{split} &\CI( |\psi_L\rangle \otimes |\psi_R\rangle,\, |\phi_L\rangle \otimes |\phi_R\rangle) \\
&\geq \min\left(\CI( |\psi_L\rangle,\, |\phi_L\rangle),\, \CI( |\psi_R\rangle,\, |\phi_R\rangle)\right ),\end{split}\end{align}
\begin{align}\begin{split} &\CD( |\psi_L\rangle \otimes |\psi_R\rangle,\, |\phi_L\rangle \otimes |\phi_R\rangle)\\ &\leq\min(\CD( |\psi_L\rangle,\, |\phi_L\rangle), \CD( |\psi_R\rangle,\, |\phi_R\rangle) )\end{split}.\end{align}
So branching is more favorable if it removes entanglement like this, because $\CI$ may grow with more long-range entanglement while $\CD$ does not.

\section{Branching versus error correction}\label{sec:error_correction}
The phrase ``good branches are the opposite of good error correcting codes" is a useful way to think about branches. Good error correcting codes protect relative-phase information between codewords from being affected by errors, so that circuits designed to get relative-phase information are unlikely end up accidentally disrupting it. Good branches allow relative phase information to be easily disrupted (in just $\CD$ gates) compared to the difficulty of getting it ($\CI$ gates). This intuition can be formalized by rewriting the conditions for error correcting codes in terms of $\CI$ and $\CD$.

B\'eny and Oreshkov gave a necessary and sufficient condition for any approximate quantum error correcting code~\cite{beny2010general_AQECC_conditions}:
\begin{equation}\label{eq:aqecc_conditon_PEEP}PE_\m^\dagger E_\n P = \lambda_{\m\n} P + PB_{\m\n}P,
\end{equation}
where $P = \sum_{\vi=1}^{2^k} |\psi_\vi\rangle\langle \psi_\vi |$ is a projector onto a subspace spanned by some subset of codewords $\{ |\psi_1\rangle, \dots, |\psi_{2^k}\rangle\}\subset 
\mathbb{C}^{2^N}$, 
$E_\m$ and $E_\n$ are error operators defining an approximately 
correctable noise channel $\mathcal{N}(\rho)=\sum_k E_k \rho E_k^\dagger=\sum_{\m\n}\lambda_{\m\n}\Tr(\rho)|\m\rangle\langle \n|$,  
while $\lambda_{\m\n}$ is any Hermitian matrix, and $B_{\m\n}$ is a matrix allowing for $\varepsilon$-approximateness of the error correcting code by a bound on the Bures distance:
$d\left(\mathcal{N}(\rho),\;\; \mathcal{N}(\rho) + \sum_{\vi\j}\Tr(\rho B_{\vi\j}) |\vi\rangle \langle \j| \right)\leq \varepsilon$. 

If this B\'eny-Oreshkov condition is met, the codeword states encode $k$ logical qubits into $N$ physical qubits while allowing for $\varepsilon$-approximate correction of any error in $\{E_\n\}$. For clarity, let us expand the projector $P$ in equation (\ref{eq:aqecc_conditon_PEEP}), giving
\begin{equation}\label{eq:aqecc_conditon_full}
\langle \psi_\vi | E_\m^\dagger E_\n  |\psi_\j\rangle   =  \lambda_{\m\n} \delta_{\vi\j} + \varepsilon_{\m\n\vi\j},
\end{equation}
where $\varepsilon_{\m\n\vi\j}=\langle \psi_\vi |B_{\m\n}  |\psi_\j\rangle$ parameterize the approximateness of the error correcting code.  If $\varepsilon_{\m\n\vi\j}=0$ for all $\m,\n,\vi,\j$, the Knill-Laflamme condition for exact error correction is recovered~\cite{knill2000theory_QECC_conditions}. 

A consequence of equation (\ref{eq:aqecc_conditon_full}) is
\begin{equation}
     \left| \langle \psi_\vi | E_\m^\dagger E_\n  |\psi_\vi\rangle  - \langle \psi_\j | E_\m^\dagger E_\n  |\psi_\j\rangle \right|  = \left|  \varepsilon_{\m\n\vi\vi} - \varepsilon_{\m\n\j\j} \right|.
\end{equation}
Let $\epsilon := \max_{\m\n\vi\j} | \varepsilon_{\m\n\vi\j}|$ and suppose that any errors $E_\n$ below some complexity $c$ are correctable. Then, for any unitary $U$ of complexity less than $2c$:
\begin{equation}\left|\langle \psi_\vi | {U} |\psi_\vi \rangle  - \langle\psi_\j | {U} |\psi_\j  \rangle\right| < 2\epsilon .\end{equation}
By the definition of distinguishability complexity (definition \ref{def:CD}), and the fact that it increases monotonically in $\Delta$, this implies
\begin{equation}\CDprox\left(|\psi_\vi\rangle,|\psi_\j\rangle, \Delta \right)>2c \, \text{ for all } \Delta \geq \epsilon.\end{equation}

Another consequence of equation (\ref{eq:aqecc_conditon_full}) is
\begin{equation} \left| \langle \psi_\vi | E_\m^\dagger E_\n  |\psi_\j\rangle\right| + \left|\langle \psi_\j | E_\m^\dagger E_\n  |\psi_\vi\rangle \right| = \left|   \varepsilon_{\m\n\vi\j} + \varepsilon_{\m\n\j\vi} \right| \text{ if } \vi\neq\j.  \end{equation}
Again let $\epsilon := \max_{\m\n\vi\j} | \varepsilon_{\m\n\vi\j}|$ and suppose that any errors $E_n$ below some complexity $c$ are correctable. Then, for any unitary $U$ of complexity less than $2c$,
\begin{equation}\left|\langle \psi_\vi | {U} |\psi_\j \rangle\right|  + \left|\langle\psi_\j | {U} |\psi_\vi  \rangle\right| < 2\epsilon \,\text{  if } \vi\neq\j .\end{equation}
By the definition of interference complexity (definition \ref{def:CI_proxy}), and the fact that it increases monotonically in $\Delta$, this implies 
\begin{equation}\CIprox\left(|\psi_\vi\rangle,|\psi_\j\rangle, \Delta \right)>2c  \, \text{ for all } \Delta \geq \epsilon\,\text{  if } \vi\neq\j .\end{equation}

\begin{figure}
    \centering\includegraphics[width=1.02\linewidth]{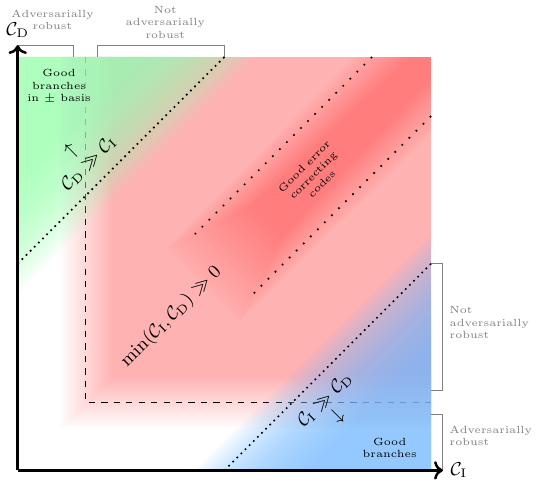}
    \caption{  Error correcting codes require $\min(\CI,\CD)\gg1$ between codewords. The best, most efficient error correcting codes occur when $\CI\approx\CD\gg1$ (assuming unbiased noise), whereas good branches only occur when $\CI\gg\CD$ (or $\CD\gg\CI$ in the $+/-$ basis).  This is the sense in which branches are the opposite of good error correcting codes. The overlapping regions where both $\min(\CI,\CD)\gg1$ and $\CI\gg\CD$ hold are ``non-adversarially robust'' regimes where it is theoretically possible to get relative-phase information with sufficiently frequent and careful quantum error correction, but where time evolution is unlikely to do this naturally.} 
    \label{fig:branches_opposite_of_good_codes}
\end{figure}

So if approximate quantum error correction is possible for all errors below some complexity $c$, then the distinguishability and interference complexities between codewords must be at least $2c$. The protection of the code is determined by $\min(\CI,\CD)$, so the most efficient codes occur when $\CI\approx\CD\gg1$ (assuming approximately balanced noise). So good branches (with $\CI\gg\CD$) are effectively the opposite of good (balanced) error-correcting codes, as illustrated in figure \ref{fig:branches_opposite_of_good_codes}.

As a result, branches are effectively the opposite of good quantum error correcting codes under our definition: they protect ``which-branch'' information, but \textit{anti-protect} relative phase information between the branches. 

For a different relationship between branching and quantum error correction, see appendix \ref{sec:branching-in-qec-environment-ancilla}, where we discuss how branches may still form between the environment and ancilla qubits in the process of quantum error correction.

\begin{figure*}
\centering
\includegraphics[width=\linewidth]{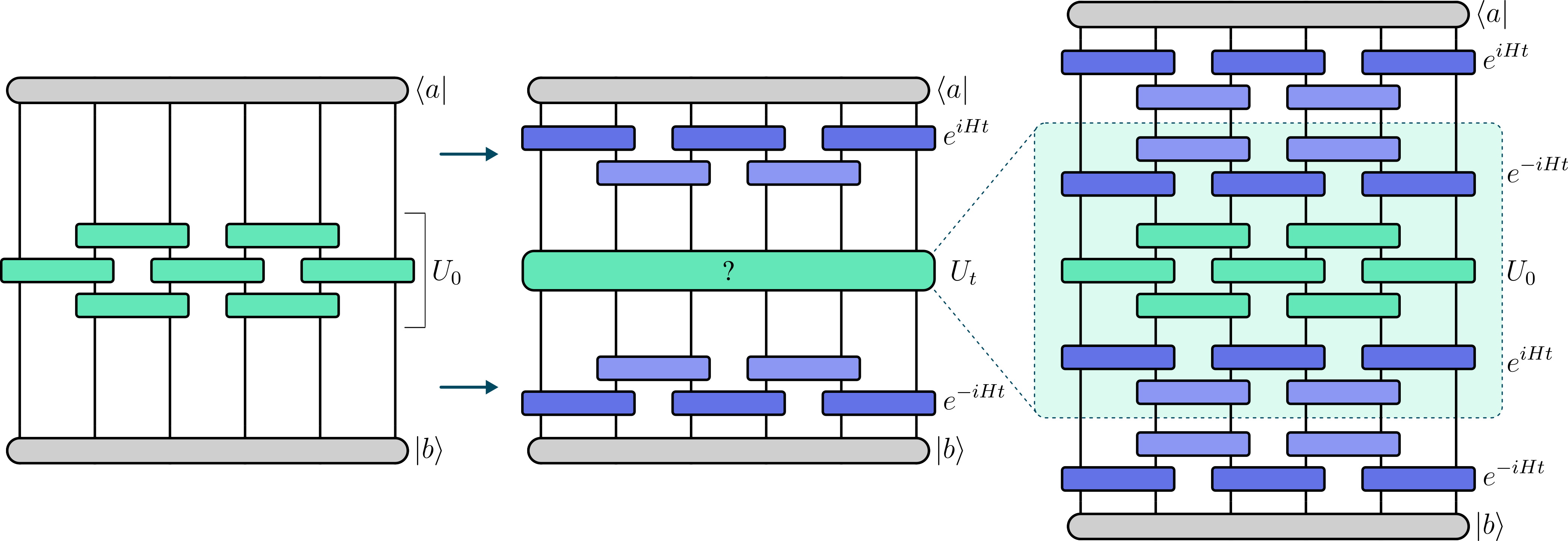}
\caption{If the initial minimal complexity unitary is $U_0$ (left), then what is the minimal complexity unitary  $U_t$ after time evolution of $|a\rangle$ and $|b\rangle$? One possibility is the concatenation of three circuits: $e^{-i{H}t}$, ${U_0}$, and $e^{i{H}t}$, though a shorter circuit may exist, especially if the initial complexity of ${U_0}$ is sufficiently small that $e^{-i{H}t}$ and $e^{i{H}t}$ can partially cancel. In general, it is conjectured that the complexity growth rate $\frac{d\C(U_t)}{dt}$ is a nearly monotonically increasing function of the complexity $\C(U_t)$, going to zero as $\C(U_t)$ goes to zero, and saturating to a constant when  $\C(U_t)$ is large, independent of $\ket{a}$ and $\ket{b}$. This implies that branches virtually only grow further apart under random circuit evolution, as we explore in section \ref{sec:time_evolution}.}\label{fig:branch_complexity_growth}
\end{figure*}

\subsection{Adversarially robust branches and rectangular surface codes}\label{sec:adversarial_robustness}
Good examples for thinking about the interplay between error correction and branching can be found in rectangular surface codes~\cite{lee2021rectangular_surface_code_toric}, such as in example \ref{sec:example:toric}, or the  $L\times l$ toric code with $L > l$, as illustrated in figure \ref{fig:toric_code}. Denote two logical codewords of a rectangular toric code as $|\bar0\rangle$ and $|\bar1\rangle$. Suppose it takes $L$ gates to swap $|\bar0\rangle$ with $|\bar1\rangle$ (going around the torus the long way) so $\CI(|\bar0\rangle, |\bar1\rangle) =L$, whereas it takes only $l$ gates to swap $\frac{|\bar0\rangle+|\bar1\rangle}{\sqrt{2}}$ with $\frac{|\bar0\rangle-|\bar1\rangle}{\sqrt{2}}$ (going around the torus the short way), so $\CD(|\bar0\rangle, |\bar1\rangle)=l$. Getting relative-phase information between $|\bar0\rangle$ and $|\bar1\rangle$ (distinguishing $\frac{|\bar0\rangle+|\bar1\rangle}{\sqrt{2}}$ from $\frac{|\bar0\rangle-|\bar1\rangle}{\sqrt{2}}$) requires going around the torus the long way, applying $\CI$ gates, whereas disrupting the relative phase requires going around the torus the short way, applying only $\CD$ gates, as shown in figure \ref{fig:toric_code}.

If $L\gg l$ and no active error correction is applied, errors are likely to cause the path around the torus to accidentally wind around the short way in the process of going around the long way, disrupting the relative-phase in the process of trying to measure it. So with no active error correction, $\CI\gg\CD$ is sufficient to make interference experiments infeasible here, even when error correcting code conditions are also satisfied (when $\min(\CI,\CD)\gg1$). Call this the regime where branches are environmentally robust (successful interference experiments are unlikely to be performed naturally) but not necessarily adversarially robust. 

It is true more generally that it takes only $O(\CD)$ uncorrected errors to disrupt the relative-phase between branches (because $\CD(|a\rangle,|b\rangle)$ gates can approximately swap $\frac{|a\rangle + |b\rangle}{\sqrt{2}}$ with $\frac{|a\rangle - |b\rangle}{\sqrt{2}}$), whereas at least $\CI$ gates are required in order to get information on the relative-phase. Additionally, if the evolution is noisy and no error correction is applied, the fidelity of the interference unitary drops exponentially with $\CI$, quickly leading to the wrong transformation~\cite{dalzell2021random_circuits_local_noise_into_global_brandao,quek2022exponentially_bounds_error_mitigation, takagi2022fundamental_limits_error_mitigation}. 
So if active error correction is not performed, $\CI\gg\CD$ is enough of a condition alone to render interference experiments getting information on the current relative-phases infeasible in general (the difficulty of getting information on \textit{future} relative phases will be considered in section \ref{sec:time_evolution}). 

Coming back to the rectangular surface code, if active error correction is applied frequently enough then the ``long'' path required to perform an interference experiment can be implemented without accidentally creating a ``short'' path which would disrupt the relative phase. When the rate of physical errors per round of error correction ($p$) is low, the rate of logical relative-phase errors ($P$) can be approximated as~\cite{lee2021rectangular_surface_code_toric}
\begin{equation}P \approx  \frac{l! L  p^{\lceil l/2 \rceil}}{\lceil l/2 \rceil!\lfloor l/2 \rfloor! } \approx L \sqrt{\frac{2l}{\pi (l+1)^2}} \left(\frac{4 l^2}{l^2-1} p \right)^{l/2+1}, \end{equation}
so to keep the rate of logical relative-phase errors above a constant when physical errors are actively corrected, $L$ must scale as $e^{c\,l\ln\left(\frac{1}{p}\right)}$ for some constant $c$. In this case interference experiments are infeasible when $\CI > e^{c\,\CD \ln\left(\frac{1}{p}\right)}$, even allowing for frequent active error correction. Although this scaling is exponential, it is always possible to satisfy for any nonzero noise rate $p$, so long as $\CD$ is sufficiently small or the total number of physical qubits is sufficiently large. Call this the regime where branches are adversarially robust.

We conjecture that this pattern holds more generally: $\CI\gg\CD$ alone is sufficient for branches to be robust in non-adversarial situations (successful interference experiments are unlikely to be performed naturally), whereas if $\CI > e^{\lambda\,\CD}$ for some constant $\lambda$ depending only on the noise rate and the frequency of active error correction then branches are adversarially robust (interference experiments are unlikely to be successful, period). We mostly focus on the non-adversarial case unless we specify otherwise, imagining that branches are simply undergoing time evolution rather than specifically directed attempts at getting relative-phase information with frequent active error correction.

\section{Time evolution}\label{sec:time_evolution}

\begin{figure}[!h]
    \centering
    \includegraphics[width=0.8\linewidth]{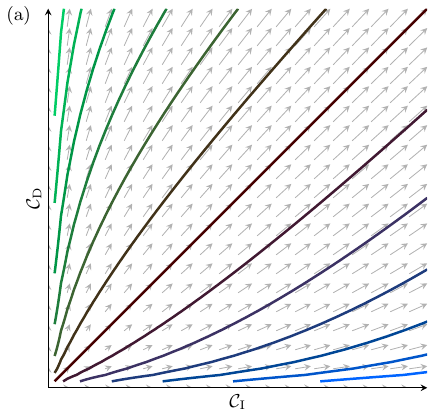}
    \qquad
    \includegraphics[width=0.8\linewidth]{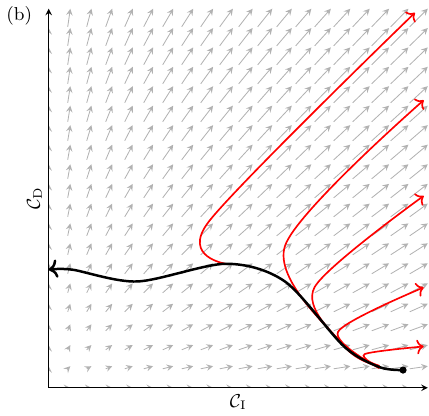}
    \caption{(a) The growth rates of complexities should be monotonically increasing functions of complexity (statistically speaking), with the same terminal growth rates. An illustrative model of this is $\frac{d }{d  t}\CI\propto\frac{\CI}{\CI+k} $ and  $\frac{d }{d  t}\CD\propto\frac{\CD}{\CD+k} $ (plotted as a vector field for $k=1$), which implies that  $\CI(t) + k\log(\CI(t)/k) \approx \CD(t) + k\log(\CD(t)/k) + b$ (plotted as flow lines for $k=1$ and varying $b$). 
    (b) The application of any good interference circuit must reduce the interference complexity to zero, as illustrated with the black path. When $\CI\gg\CD$, small deviations quickly lead $\CI$ and $(\CI-\CD)$ to increase rather than decrease, unless active error correction is regularly applied~\cite{halpern2021resource_theory_complexity} (which is also infeasible if $\CI> e^{\lambda\CD}$, as argued in section \ref{sec:error_correction}). }
    \label{fig:growth_flow_with_time}
\end{figure}

For a good branch decomposition, feasible experiments shouldn't be able to differentiate the full quantum state $\rho$ from the diagonal density matrix of branches $\rho_{\text{diag}}$, even under very long time evolution $t\gg\CI$. We want to show that if $\CI - \CD$ is large at one time, then under reasonable assumptions $\CI$ will grow with time at least as fast as the growth of $\CD$. If this is true, then branches will stay branches over time (under definition \ref{def:valid_branch}), only growing ``further apart'' in complexity space rather than coming back together. 

We introduce two arguments that branches virtually only grow further apart (when we say ``virtually'', we mean it in the same sense as ``the second law of thermodynamics is virtually never violated''). The first argument (presented in this section) is based on conjectures which have their strongest footing in random circuits, while the second (presented in section \ref{sec:conserved_quantities_eth_thermalization}) is based on the presence of conserved quantities. 

Here we will argue that the growth rates $\frac{d \CI}{dt}$ and $\frac{d \CD}{dt}$ are virtually monotonically increasing functions of $\CI$ and $\CD$ (so long as neither complexity is exponentially large), with the same growth dynamics for random circuits, and the same terminal growth rates (ignoring conserved quantities). So if $\CI \gg \CD$, then $\frac{d }{d  t}\CI >\frac{d }{d  t}\CD$ with high probability, implying that branches will only grow further apart until $\CI$ is exponentially large.

Suppose we know the number of gates $\CI(\ket{a}, \ket{b}, \epsilon)$ in the initial minimum-complexity interference circuit $U_0$ satisfying $|\langle a| {U_0} |b\rangle|  + |\langle b| {U_0} |a \rangle| \geq 2\epsilon $ (definition \ref{def:CI_proxy}). We want to estimate the number of gates $\CI(\ket{a(t)}, \ket{b(t)}, \epsilon)$ in the minimum-complexity unitary $U_t$ satisfying this at later times: 
\begin{equation}\label{eq:CI_time_evo_condition}
    \left|\langle a|e^{i{H}t} {U_t} e^{-i{H}t} |b\rangle\right|  + \left|\langle b|e^{i{H}t} {U_t} e^{-i{H}t}|a \rangle\right| \geq 2\epsilon .
\end{equation}
One choice of $U_t$ which would satisfy equation (\ref{eq:CI_time_evo_condition}) is the negative time evolution of the original unitary: $U_t \approx e^{-i{H}t} {U_0} e^{i{H}t}$. It is possible there may be a very different $U_t$ also satisfying equation (\ref{eq:CI_time_evo_condition}) with a complexity much less than $\C(e^{-i{H}t} {U_0} e^{i{H}t})$.  If $e^{i{H}t}$ were a random circuit this would be extremely unlikely, by the conjecture that ``collisions'' between random circuits of sub-exponential size are extremely rare (supposing $t$ and $\CI$ are not yet exponentially large in the system size)~\cite{brown2018second, brandao2021models, haferkamp2022linear}. We will analyze the relevant differences between Hamiltonian and random-circuit evolution in section \ref{sec:conserved_quantities_eth_thermalization}, but if we neglect these differences for the moment we get $\C(U_t) \approx \C(e^{-i{H}t} {U_0} e^{i{H}t})$.


So what is the complexity of the operator $e^{-i{H}t} {U_0} e^{i{H}t}$, and how does it compare to $\C(U_0)$ (which is constant) and $\C(e^{i{H}t})$ (which generally grows linearly in time~\cite{brown2018second, brandao2021models, haferkamp2022linear, eisert2021entangling_power_complexity})? Figure \ref{fig:branch_complexity_growth} illustrates this question. It has been studied in the context of black holes, where $\C(e^{-i{H}t} {U_0} e^{i{H}t})$ is known as the ``precursor complexity''~\cite{susskind2014switchbacks,brown2017quantum_complexity_curvature}.  We might choose to view the initial circuit $U_0$ as some ``perturbation'' operator, in which case $\C(e^{-i{H}t} {U_0} e^{i{H}t})$ is a measure of how much that perturbation has spread under the dynamics of the system.  Indeed, a very similar quantity  $\C(e^{-i(H+\delta H)t} e^{i{H}t})$ has been proposed as a diagnostic of quantum chaos~\cite{ali2020chaos_complexity}. 

Consider the growth of the precursor complexity $\C(e^{-i{H}t} {U_0} e^{i{H}t})$ over time.  
If $\C({U_0})$ extremely small (a single gate, say) and $t$ is small, then there will exist a much simpler circuit implementing the same transformation as $e^{-i{H}t} {U_0} e^{i{H}t}$ than applying $e^{-i{H}t}$, ${U_0}$, and $e^{i{H}t}$ sequentially, because cancellations occur between $e^{-i{H}t}$ and $e^{i{H}t}$. In this case, $\C(e^{-i{H}t} {U_0} e^{i{H}t}) $ will be small ($\ll \C({U_0}) + 2\C(e^{i{H}t})$) and grow slowly with $t$. 
In the limit of large $\C({U_0})$ or large $t$ however, essentially no cancellations are conjectured to occur between $e^{-i{H}t}$ and $e^{i{H}t}$ (for chaotic Hamiltonians or random circuits), and $\C(e^{-i{H}t} {U_0} e^{i{H}t}) \approx \C({U_0}) + 2\C(e^{i{H}t})-c$, growing linearly with time, with a constant offset $c$ (called a ``switchback'') caused by the initial delay in linear growth during the scrambling time~\cite{susskind2014switchbacks,brown2017quantum_complexity_curvature, ali2020chaos_complexity}. This also assumes that cancellations between $U_0$ and $e^{i{H}t}$ and between $e^{-i{H}t}$ and $U_0$ are rare, which is justified for the vast majority of time evolution operators not specifically tuned to decrease the complexity of $U_0$~\cite{brown2018second, brandao2021models, haferkamp2022linear, eisert2021entangling_power_complexity}, unless $U_0$ commutes with $H$, as we will investigate in section \ref{sec:conserved_quantities_eth_thermalization}. Even if these time evolution operators are specifically tuned to decrease complexity or commute with $U_0$, small perturbations in them or in  $U_0$ will rapidly lead back to increasing complexities~\cite{halpern2021resource_theory_complexity, brown2017quantum_complexity_curvature}. 

\newcommand{\approptoinn}[2]{\mathrel{\vcenter{
  \offinterlineskip\halign{\hfil$##$\cr
    #1\propto\cr\noalign{\kern2pt}#1\sim\cr\noalign{\kern-2pt}}}}}

\newcommand{\appropto}{\mathpalette\approptoinn\relax}

Overall, the complexity $\CI(\ket{a(t)},\ \ket{b(t)},\  \epsilon)$ of the interference unitary $U_t$ should grow at the same speed as the precursor complexity $\C(e^{-iHt}U_0e^{iHt})$: linear in time $\propto N t$ when $\C(U_0)\gg N$ (where $N$ is the system size), and slower when $\C(U_0)< N$, going to zero as $\C(U_0)\rightarrow0$. Although the complexity growth initially depends on the initial circuit $U_0$, it will eventually plateau to the same linear growth rate regardless of $U_0$ or $\ket{a}$ or $\ket{b}$ (dependent only on $H$), and at intermediate times the complexity growth should be determined largely monotonically by $\C(U_0)$, rather than the specifics of $U_0$ itself (assuming the evolution is sufficiently similar to random-circuit evolution). Importantly, all of the above arguments hold equally for distinguishability complexity as well as interference complexity. So if $\CI(t=0) \gg \CD(t=0)$, then $\frac{d }{d  t}\CI(t=0) >\frac{d }{d  t}\CD(t=0)$ and $\CI(t) \gg \CD(t)$ with high probability. Therefore good branch decompositions (definition \ref{def:valid_branch}) should remain good branch decompositions with time, at least until the interference complexity is exponentially large in the system size (if random circuits are a relevant proxy for time-evolution). When we move away from random circuits in section \ref{sec:conserved_quantities_eth_thermalization} to systems with conserved quantities, we argue that this only favors the growth of $\CI$ relative to $\CD$, further increasing the extent to which branches grow apart with time. 

Although this is all we need to show that branches are unlikely to come back together under random-circuit time evolution, it helps to see more concrete models for this growth of $\CI$ and $\CD$. 
The precise nature of complexity growth will depend on the scrambling dynamics of the Hamiltonian (and therefore on the dimensionality of the system), but 
an illustrative model of complexity growth is shown in figure \ref{fig:growth_flow_with_time}. This model captures the relevant dynamics of polynomial growth at early times (as the influence of the initial circuit $U_0$ expands locally), saturating to maximal linear growth at longer times (once the perturbation has been well spread across the system and cancellations no longer occur between $e^{-iHt}$ and $e^{iHt}$). Previous explicit models of precursor complexity have also been considered in the context of systems with spatially nonlocal ``fast scrambling'' interactions, which are similar but have exponential growth at early times~\cite{susskind2014switchbacks}.

These models illustrate the arguments that $\CI$ and $\CD$ will both have growth rates statistically monotonically increasing in the same way with initial complexity, converging to the same maximal rates of growth at long times (modulo considerations in section \ref{sec:conserved_quantities_eth_thermalization}). As such, if the difference between the interference complexity and distinguishability complexity is initially large, it should almost always grow with time. Even in the case of a time evolution operator specifically constructed to reduce the interference complexity, if $\CI\gg\CD$ then small perturbations will quickly lead the interference complexity to increase again (and at a faster rate than $\CD$) unless active error correction is regularly applied~\cite{brown2017quantum_complexity_curvature}. 

Although specific models are useful for illustrative purposes, all we really require is that the growth rates $\frac{d \CI}{dt}$ and $\frac{d \CD}{dt}$ are similar monotonically increasing functions of $\CI$ and $\CD$ on average when $\CI$ and $\CD$ are not exponentially large. 
Our main assumptions here are that evolution occurs under at least a slightly scrambling Hamiltonian (or with the presence of slightly scrambling noise), that the Hamiltonian lacks enough conserved quantities to make a small interference circuit which commutes with it (relaxed in section \ref{sec:conserved_quantities_eth_thermalization}), that $\CI$ is much smaller than the maximum (exponential in system size) complexity, and that the ``second law of quantum complexity'' conjectures of Brown and Susskind are valid~\cite{brown2017quantum_complexity_curvature, brown2018second}. See appendix \ref{sec:irreversibility} for a simpler link between branching and irreversibility, independent of these conjectures.

\section{Thermalization and conserved quantities}\label{sec:conserved_quantities_eth_thermalization}
The Eigenstate Thermalization Hypothesis (ETH) conditions can be formulated as 
\begin{equation} \label{eq:eth1}
\left|\langle E_\k|O|E_\k\rangle - \langle E_{\k+1}|O|E_{\k+1}\rangle\right|\leq e^{-c_1N}
\end{equation}
and
\begin{equation} \label{eq:eth2}
|\langle E_\k|O|E_\m\rangle | \leq e^{-c_2N},\  k\neq m
\end{equation}
where $|E_\k\rangle$ are any energy eigenstates from the bulk of the spectrum of a ``chaotic'' Hamiltonian, $O$ is any local observable, $N$ is the system size, and $c_1$ and $c_2$ are constants~\cite{srednicki1994chaos_thermalization_eth,brandao2019quantum_eth_error_correction_translational_invariance}. The first equation implies that nearby energy eigenstates in the bulk of the spectrum are indistinguishable to local expectation values, while the second implies that fluctuations of local expectation values around the long-time average are small~\cite{brandao2019quantum_eth_error_correction_translational_invariance}. 

These equations bear striking resemblances to our definitions for $\CDprox$ and $\CIprox$ respectively (definitions \ref{def:CI_proxy} and \ref{def:CD_proxy}). 
If the ETH holds (equations \ref{eq:eth1} and \ref{eq:eth2}) for all operators ${O}$ below some complexity $C$, then the distinguishability complexity between nearby eigenstates $\CD(|E_\k\rangle,\, |E_{\k+1}\rangle,\, e^{-c_1N}) $ must be at least $C $, and the interference complexity between any two different eigenstates $\CI(|E_\k\rangle,\, |E_{\m}\rangle,\, e^{-c_2N}) \ (\k \neq \m)$ must also be at least $C$.
Alternatively, a conceptually similar version of the ETH can be formulated by requiring that the distinguishability complexity between nearby energy eigenstates ($\CD(|E_\k\rangle,\, |E_{\k+1}\rangle,\, \Delta)$) scales exponentially with the system size, as does the interference complexity between any two different eigenstates ($\CI(|E_\k\rangle,\, |E_{\m}\rangle,\, \Delta)$).\footnote{In order to translate between these two formulations of the ETH, it would be useful to have upper bounds on how complexity $\C(\ket{a},\ket{b}, \Delta)$ can increase as the tolerance $\Delta$ is decreased.} This may be a useful formulation for generalizing the spirit of the original ETH to make statements about more complex observables. 

The resulting condition that $\CI\sim\CD\gg1$ between nearby ETH eigenstates implies that they make good error-correcting codes (see section \ref{sec:error_correction}), as has already been noted~\cite{brandao2019quantum_eth_error_correction_translational_invariance,bao2019eigenstate_eth_error_correction}. A connection between the ETH conditions and exponential distinguishability complexity of nearby eigenstates has also already been established by a link to Grover's algorithm~\cite{bao2021microstate}.

As we consider ETH eigenstates of very different energies however, $\CI$ will remain large while $\CD$ may reduce (as equation \ref{eq:eth1} is no longer relevant): it is easier to locally distinguish eigenstates of very different energies, but not easy to swap them. This points to the formation of branches in chaotic thermalizing systems: whenever there is a superposition of quite different energy scales we can expect multiple branches - one for each energy scale, as shown in figure \ref{fig:energy-branches} (a). If $\CI$ continues to grow while $\CD$ is limited, branching would gradually occur between finer and finer energy scales as shown in figure \ref{fig:energy-branches} (b).

One might worry that because most energy eigenstates are highly entangled~\cite{bianchi2022volume_law_most_states}, there will be a trade-off between branching which reduces energy variance like this and branching which reduces entanglement as seen in section \ref{sec:example:entangled}. However recent work has shown that there is generally a lot of room for energy variance to decrease down to at least $\delta^2\propto1/\log(N)$ without significantly increasing entanglement above $\log(N)$~\cite{rai2023matrix_approximations_low_energy_variance}, so this trade-off may not kick in strongly until relatively late times. 
{%
\begin{figure}
    \centering
\includegraphics[width=\linewidth]{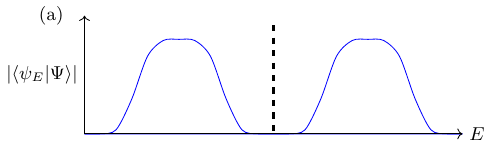}
\includegraphics[width=\linewidth]{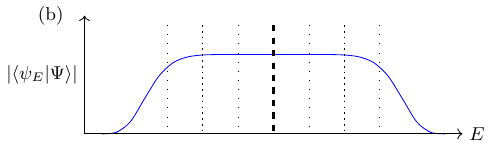}
    \caption{(a) When a many-body quantum system satisfying the eigenstate thermalization hypothesis is in a superposition of well-separated energy regions, we expect these to form good branches, as denoted by the dashed line. (b) If the occupation is not separated across the energy spectrum, there will be occupied eigenstates which are hard to distinguish around any potential cut for branching, increasing $\CD$. Nonetheless, branches may still form over time as $\CI$ increases. In order to minimize the relative weight of the components in each branch which are hard to distinguish, these branches should form in a pattern like the one shown, starting with the thick dashed cut, then progressing to the thinner cuts as $\CI$ continues to grow.   }
    \label{fig:energy-branches}
\end{figure}
}

This is a specific example of a more general effect: symmetries facilitate branching, because of fundamental trade-offs between symmetries and error correction
~\cite{eastin2009restrictions_eastin_knill,faist2020continuous_symmetries_error_correction,kubica2021using_simple_proof_approx_eastin_knill,tajima2021universal_limitation_symmetry,liu2021quantum_error_correction_symmetries,liu2021approximate_symmetry_quantum_error_correction,zhou2021new_perspectives_covariant_error_correction,yang2022optimal_error_correction_reference_frames,tajima2022universal_trade_symmetry}. 

In the context of branching, symmetries make it easier to distinguish states by distinguishing the relevant conserved quantities, reducing $\CD$. Additionally, circuits able to measure these conserved quantities are more likely to commute with the Hamiltonian, greatly slowing or stopping the growth of $\CD$ with time. 

To see this, recall that the distinguishability complexity at some time $t$, $\CDprox(\ket{a(t)}, \ket{b(t)}, \Delta)$, is the minimum number of gates in a circuit $U_\text{D}$ satisfying
\begin{equation}\label{eq:CD_time_evo}
    \left|\langle a|e^{i{H}t} {U_\text{D}} e^{-i{H}t} |a\rangle  - \langle b|e^{i{H}t} {U_\text{D}} e^{-i{H}t}|b \rangle\right| \geq 2\Delta.
\end{equation}
If symmetries are present (including but not limited to energy conservation), then there is some set of 
unitaries $\{\mathcal{U}_i\}$ (e.g. the symmetry operators) which commute with the Hamiltonian: $H\mathcal{U}_i=\mathcal{U}_iH$. For example, if there is some Hermitian symmetry generator $G$, then the unitary operators  $\mathcal{U}(\theta) = e^{iG\theta}$ would all commute with the Hamiltonian, and small $\theta$ operators would likely correspond to relatively low-depth circuits. If a circuit for performing a relevant operation (eg. interference, or distinguishability) could be built entirely from these symmetry operations, then this circuit could be used forever and the complexity would not increase beyond this level, because $e^{i{H}t}$ and $e^{-i{H}t}$ would commute past the circuit and cancel.

Furthermore, if we suppose that $\ket{a}$ and $\ket{b}$ are approximate eigenstates of some symmetry operator $\mathcal{U}$, then $\mathcal{U}\ket{a}\approx e^{i\theta}\ket{a}$ and $\mathcal{U}\ket{b}\approx e^{i\phi}\ket{b}$ for some phases $\theta$ and $\phi$.  Using this symmetry operator for $U_\text{D}$ reduces equation \ref{eq:CD_time_evo} to
\begin{equation}
    \left|e^{i\theta} - e^{i\phi} \right| \geq 2\Delta,
\end{equation}
so $\CD \leq O[\C(\mathcal{U})]$ for all times if $\theta \approx \phi \pm \pi$: the symmetry operator can prevent the growth of distinguishability complexity between its eigenstates. The same is not true for the interference complexity. Recall that $\CIprox(\ket{a(t)}, \ket{b(t)}, \Delta)$ is the minimum number of gates in a circuit $U_\text{I}$ satisfying
\begin{equation}
    \left|\langle a|e^{i{H}t} {U_\text{I}} e^{-i{H}t} |b\rangle\right|  + \left|\langle b|e^{i{H}t} {U_\text{I}} e^{-i{H}t}|a \rangle\right| \geq 2\Delta,
\end{equation}
which only reduces under the symmetry operator $\mathcal{U}$ to 
\begin{equation}
    \left|e^{i\phi}\braket{a}{b}\right|  + \left|e^{i\theta} \braket{b}{a}\right| \geq 2\Delta,
\end{equation}
so the symmetry operator is essentially useless for performing interference experiments between its eigenstates in different symmetry sectors, because it cannot map between them. The interference complexity between approximate symmetry eigenstates can therefore keep growing with time while the distinguishability complexity plateaus at $O[\C(\mathcal{U})]$.

{Practically, two things are important for determining how symmetries affect the growth rates of $\CI$ and $\CD$: the complexity of the symmetry operators themselves, and their usefulness for constructing interfering or distinguishing circuits. If the symmetry operators are all highly complex, then they cannot be used in any lower complexity circuits, and so their effects on complexities won't show up until then. Relatedly, if the symmetry operators are too restricted to be useful for constructing interference or distinguishablility circuits between the branches, then they will not affect these complexity growth dynamics either. Therefore this effect will be strongest when there is a broad set of low-complexity unitaries which commute with the Hamiltonian. }

This kind of mechanism may give an alternate way of explaining the appearance of superselection rules, complementary to einselection in the environmentally-induced decoherence or decoherent histories formalisms~\cite{zurek2003decoherence, giulini2003superselection}.  

The fact that some states remain easy to distinguish over time makes intuitive sense. After all, it usually seems easy to distinguish a system at one energy from the system at a very different energy, even long after thermalization times where most other information would be scrambled - it's not too hard to measure the temperature, after all.  Likewise for properties which aren't perfectly conserved, but are effectively conserved on relevant timescales, like emergent hydrodynamic variables (densities which vary relatively slowly in large enough sub-volumes~\cite{halliwell2010macroscopic}),  or whether a cat is alive or dead (hopefully). This may significantly alter figure \ref{fig:growth_flow_with_time} to favor the growth of $\CI$ over $\CD$, and cause adversarially robust branches to remain adversarially robust over time (definition \ref{def:great_branch}), rather than transitioning to merely ``good'' branches (definition \ref{def:valid_branch}). 

This also bears on whether the difference $\CI-\CD$ is really the most relevant ``branchiness'' quantity as postulated in definition \ref{def:valid_branch}. Might other measures such as the ratio $\CI/\CD$ also be important? The previous sections have not supported this. Defining branches by anything other than the difference between $\CI$ and $\CD$ would tilt the lines in figure \ref{fig:branches_opposite_of_good_codes} away from 45\textdegree, meaning that branches would not necessarily stay branches under the model of complexity growth in figure \ref{fig:growth_flow_with_time} (a), where the long-time growth rates of $\CI$ and $\CD$ are the same. Under this random-circuit inspired model, $\CI-\CD$ seems the best ``branchiness'' measure. However the conserved quantities present in Hamiltonian evolution seem to favor the growth of $\CI$ over $\CD$, perhaps allowing the ratio $\CI/\CD$ to also be a useful ``branchiness'' measure in these systems. 

\section{Lack of a unique preferred basis}\label{sec:uniqueness}
Our definition of a good branch decomposition (definition \ref{def:valid_branch}) is not intended to single out a unique ``correct'' way of decomposing a wavefunction into branches: merely to categorize the space of possible decompositions by whether they make good or bad branches. Still, we initially hoped that different but equally valid branch decompositions of the same state would be ``compatible'', in the sense that there is some joint decomposition which approximately incorporates them (as in appendix \ref{sec:three_branch_compatibility}, for example). 

However compatibility between decompositions is not true in general. Consider the state 
\begin{equation}\label{eq:uniqueness_counterexample}
\begin{split}
\ket{\Psi} &= \frac{\ket{0}\ket{\eta_0}}{\sqrt{2}} +\frac{\ket{1}\ket{\eta_1}}{\sqrt{2}}\\  &= \frac{\ket{+}\ket{\eta_+}}{\sqrt{2}} + \frac{\ket{-}\ket{\eta_-}}{\sqrt{2}},
\end{split}
\end{equation}
where $\ket{+} = \frac{\ket{0}+\ket{1}}{\sqrt{2}}$, $\ket{-} = \frac{\ket{0}-\ket{1}}{\sqrt{2}}$, $\ket{\eta_+} = \frac{\ket{\eta_0}+\ket{\eta_1}}{\sqrt{2}}$, and $\ket{\eta_-} = \frac{\ket{\eta_0}-\ket{\eta_1}}{\sqrt{2}}$. A state like this could be generated by starting with a bell-pair, then keeping one half of the bell pair isolated while letting the other half become coupled to an environment by deep random circuit evolution.  If $\CI(\ket{\eta_0},\ket{\eta_1})\approx\CI(\ket{\eta_+},\ket{\eta_-}) \gg 1$, then both $\left[\frac{\ket{0}\ket{\eta_0}}{\sqrt{2}}, \frac{\ket{1}\ket{\eta_1}}{\sqrt{2}}\right]$ and $\left[\frac{\ket{+}\ket{\eta_+}}{\sqrt{2}} , \frac{\ket{-}\ket{\eta_-}}{\sqrt{2}}\right]$ are good branch decompositions of $\ket{\Psi}$ under our definition.

Clearly, our definition of good branches does not always pick out a unique preferred basis. This apparent problem is related to the fact that our definition does not require information to be redundantly recorded, as we consider in appendix \ref{sec:robustness}. However this is not necessarily a problem. For example, Riedel has recently argued that branches which require redundant records are unlikely to be able to absorb enough entanglement to account for classical reality~\cite{riedel2024generalizing}. More generally, he argues against definitions of branching which require a unique preferred basis, in favor of ``generalized'' branches which merely pick out preferred subspaces~\cite{riedel2024generalizing}. Still, we leave it to future work to determine whether or not our definition reliably picks out consistent preferred subspaces.

\section{Summary: treating pure states as mixed states}\label{sec:mixed_states}
In summary, we wanted a way of defining good decompositions of a state $|\Psi\rangle$ into a sum over branches $|\Psi\rangle = \sum_\vi  e^{i\theta_\vi}\sqrt{p_\vi} | \psi_\vi \rangle$ such that the results of any local operators after natural time evolution can be calculated using a classical probability distribution over the branches $\rho_{\text{diag}} = \sum_{\vi } p_\vi |\psi_\vi \rangle\langle\psi_\vi |$ rather than the pure density matrix $\rho_{} \coloneqq |\Psi\rangle \langle\Psi| = \sum_{\vi \j }e^{i\left(\theta_{\vi}-\theta_{\j}\right)} \sqrt{p_i p_j} |\psi_\vi \rangle\langle\psi_\j |$.  By ``natural'' time evolution, we mean the result of any local Hamiltonian or circuit evolution, including specifically directed time-dependent evolution, so long as Maxwell's demon is not present in the form of frequent and precisely conditioned active error-correction operations reducing the interference complexity.  

Given $\rho$ or $\rho_\text{diag}$, the difference in the probability of any outcome $m$ after applying any circuit $U$ is 
\begin{equation}
\begin{split}
&\left|P\left(m\,|\,U,\,\rho\right) - P\left(m\,|\,U,\,\rho_\text{diag}\right)\right|\\
\leq \ &  \sum_{\vi>\j} \sqrt{p_\vi p_\j} \left|
P\left(m\,|\,U,\ket{+_{\vi\j}}\right) 
- P\left(m\,|\,U,\ket{-_{\vi\j}} \right)\right|,
\end{split}
\end{equation}
where $\ket{\pm_{\vi\j}} = \frac{e^{i\theta_\vi}\ket{\psi_\vi} \pm e^{i\theta_\j}\ket{\psi_\j}}{\sqrt{2}}$. So if discriminating between $\ket{+_{\vi\j}}$ and $\ket{-_{\vi\j}}$ infeasible for all $\vi\neq \j$ (or equivalently, if detecting interference between $\ket{\psi_\vi}$ and $\ket{\psi_\j}$ infeasible for all $\vi\neq \j$), then so is discriminating between $\rho$ and $\rho_\text{diag}$. If the interference complexities $\CI(\ket{\psi_\vi},\ket{\psi_\j}, \Delta)$ between branches are all at least some value $C$, then it must take at least $C$ gates for the the probability of any outcome to be different between $\rho_\text{diag}$  and $\rho(\theta_1,\theta_2,\dots)$ at any relative-phases $\theta_1,\theta_2,\dots$ by at least $\sum_{\vi>\j}\sqrt{p_\vi p_\j} \Delta$. 

Of course, just because an operation requires many gates doesn't mean it's infeasible. To that end, our definition compares the interference complexity $\CI$ to the distinguishability complexity $\CD$.  When $\CI-\CD$ is large between every pair of branches, it takes very many gates to get information on the relative-phases differentiating $\rho$ from $\rho_{\text{diag}}$, but many fewer gates to disrupt this relative-phase information. We argued in section \ref{sec:adversarial_robustness} that without active error-correction or near-perfect control over the Hamiltonian, near-perfect isolation from noise, and near-perfect knowledge of the initial state, this leads to unrecoverable disruption of the current relative-phases in any attempt to measure them.

However to replace $\rho$ with $\rho_{\text{diag}}$, we need to be sure that we will not be able to get information on the \textit{future} relative phases either: $\CI-\CD$ must continue to grow under time evolution. We argued in section \ref{sec:time_evolution} that complexity growth rates are monotonically increasing functions of complexity (statistically speaking, so long as complexity is not exponentially large), so if $\CI-\CD$ is large at one time, then this should continue to hold at later times under natural or random circuit time-evolution until $\CI$ is exponentially large.  

Overall, if you are given a pure state $|\Psi\rangle$ and you can find a decomposition $\sum_\vi c_\vi |\psi_\vi\rangle$ satisfying definition \ref{def:valid_branch}, then you may treat the pure state $\rho=|\Psi\rangle \langle\Psi|$ as the mixed state $\rho_{\text{diag}} = \sum_{\vi } |c_\vi |^2 |\psi_\vi \rangle\langle\psi_\vi |$, so long as you aren't worried about some sort of Maxwell's demon reducing complexity. 

Because identifying branches lets you replace them with a diagonal density matrix (a classical probability distribution over the branches), the branching of branches (or of mixed states) can be performed independently for each branch (or for each diagonal term in the density matrix), simply growing the number terms in the classical probability distribution.

\section{Conclusion}\label{sec:conclusion}
If you're given a pure state $|\Psi\rangle$ and a proposed branch decomposition $|\Psi\rangle=\sum_\vi c_\vi |\psi_\vi\rangle$, you can calculate the interference complexity $\CI(|\psi_\vi\rangle,\, |\psi_\j\rangle,\, \epsilon)$ (definition \ref{def:CI}) and distinguishability complexity $\CD(|\psi_\vi\rangle,\, |\psi_\j\rangle,\, 1-\epsilon)$ (definition \ref{def:CD})  between each of the terms, and then if each difference $\CI-\CD$ is sufficiently large, you can call the terms ``good branches'' (definition \ref{def:valid_branch}).
We have argued that such branches:
\begin{itemize}
    \item Allow the full state to be replaced with a probability distribution over the branches (a mixed state) for the purposes of natural time evolution and local measurements, so that results can be calculated separately for each branch and added probabilistically (section \ref{sec:mixed_states})
    \item Very probably do not recombine (in fact only getting better separated) for exponentially long times under natural time evolution ( $\CI$ very probably grows faster than $\CD$ if $\CI$ is initially much larger than $\CD$) (section \ref{sec:time_evolution})
    \item Imply effective irreversibility when they form from a product state (appendix \ref{sec:irreversibility})
    \item Occur in isolated systems regardless of what is happening in other isolated systems (section \ref{sec:example:separable})
    \item  Tend to absorb spatial entanglement over time (section \ref{sec:example:entangled})
    \item Are strengthened by the presence of conserved quantities (section \ref{sec:conserved_quantities_eth_thermalization})
    \item Are effectively the opposite of good error correcting codes (section \ref{sec:error_correction})
    \item Allow the results of adversarial time evolution with actively error-corrected measurements to be calculated separately for each branch and added probabilistically, so long as $\CI>\exp(\lambda \CD)$ (section \ref{sec:adversarial_robustness})
\end{itemize}

In short, branches occur when relative phase information between terms in a superposition becomes much more scrambled than information distinguishing the terms. Naturally this does not happen at a single precise moment; branches become better separated over time and in the large-$N$ limit, where $\CI$ has a chance to grow much larger than $\CD$. 

We conjecture that branch formation is a ubiquitous process in nature, occurring generically in the time evolution of many-body quantum systems. For example, we expect branches to form in quenches of chaotic systems with broad energy occupations (or broad occupations across other symmetry sectors, or across hydrodynamic variables), such as those satisfying the eigenstate thermalization hypothesis. These branches should then branch continually, leading to exponential proliferation of branches in a process accounting for much of the growth of spatial entanglement with time in quantum systems. We anticipate that this formation of branches consistent with the eigenstate thermalization hypothesis will help to explain puzzles in quantum thermalization. 

This research also highlights several questions which may now be answerable. For instance, exactly how generic is branch formation in various systems, and how do the timescales for branch formation compare to the timescales of thermalization and entanglement growth? What happens after exponentially long times, when complexities become near maximal? 

These questions and conjectures can be investigated by further theoretical investigations (such as by exploiting connections to error-correction or the eigenstate thermalization hypothesis), or by directly looking for the formation of branches in various numerical time evolution simulations. We are currently looking for this numerical evidence, by developing algorithms to find branch decompositions of tensor-network states~\cite{taylor2023branches_numerics_in_preparation}.
In addition to the direct discoveries this could facilitate, such an algorithm may also open the door to longer time evolution simulations by sampling over the branches, as proposed by Riedel~\cite{riedel2017classical}. 

Moreover, we expect our proposed definition of branches to be useful for concrete theoretical, numerical, and experimental tests of the foundations of quantum mechanics. A calculable notion of ``branchiness'' such as $\CI-\CD$ could provide a metric for experimental tests probing the boundaries of the quantum to classical transition, such as tests for violations of ``local friendliness''~\cite{bong2020strong_no_go_wigners_friend, wiseman2022thoughtful_local_friendliness}. 

Our definition also opens the door to theoretical or numerical falsifications of the many-worlds interpretation. If classical reality is characterized by a lack of spatial entanglement and superposition, then branches can only correspond to classical reality if they form rapidly enough in the relevant systems to absorb the growth of spatial entanglement with time, so that each branch has relatively little bipartite entanglement entropy, as pointed out by Riedel.\footnote{personal communication} If branches do not absorb enough entanglement to account for classical reality, then either a new definition of branches is required, or the many-worlds interpretation is false.
 
Naturally, there are several directions in which our definitions could be extended. Firstly, we could require that branch-distinguishing information be redundantly accessible, so that branches are robust to spatial coarse-grainings, as we discuss in appendix \ref{sec:robustness}. Secondly, our definitions could be extended to use Nielsen's geometric complexity, allowing for a geometric picture of branches and a more precise handling of continuous time evolution~\cite{nielsen2006quantum}. Thirdly, a distinction could be made between circuit width and circuit depth in complexities, which might enable a more complex treatment of how errors propagate to disrupt the relative-phase information. Finally, alternatives to $\CI-\CD$ as a ``branchiness'' measure could be investigated, to determine whether this is really the most relevant quantity in all natural situations, even in the presence of conserved quantities. Connections to other definitions could also be explored, such as Weingarten's~\cite{weingarten2022macroscopic}, which might allow implications to be ported across in both directions, and shed light on whether branches as we have defined them are effectively unique.

Overall, we see our work as a generalization of the basic ideas of environmentally-induced decoherence to situations with no clear system/ environment split. Rather than defining some parts of the state to be the system and others to be the environment, one may look directly for structures in the full wavefunction which correspond to the emergence of effective decoherence. \\

\section*{Acknowledgements}
We thank C. Jess Riedel and Daniel Ranard for insightful discussions about our definition, and for sharing an unpublished work from themselves, Curt von Keyserlingk, and Markus Hauru. In particular, Jess Riedel initially suggested considering error correction and quantum Darwinism in motivating the importance of distinguishability complexity, and otherwise provided much valuable feedback. Thanks also to Eleanor Rieffel, Eric Cavalcanti, Gerard Milburn, Howard Wiseman, Nora Tischler, Tom Stace, and Will Zeng for discussions and feedback, to ESI Vienna and the organisers of the \href{https://www.esi.ac.at/events/e424/}{2022 Tensor Networks programme}, and to the organisers participants and sponsors of the \href{https://www.wignersfriends.com/experiment-workshop}{2023 Wigner's Friends Workshop}. J.K.T~is supported by an Australian Government Research Training Program (RTP) Scholarship. I.P.M.~acknowledges funding from the National Science and Technology Council (NSTC) Grant No.~112-2811-M-007-044 and 113-2112-M-007-MY2, and from the Australian Research Council (ARC) Discovery Project Grants No.\ DP190101515 and DP200103760.

\bibliographystyle{unsrtnat}
\bibliography{branches}

\appendix

\section{Branching outside the system in quantum error correction}\label{sec:branching-in-qec-environment-ancilla}
We have seen in section \ref{sec:error_correction} that good codewords and good branches are effectively opposites, but error correction requires more than just choosing good encodings. Errors must actually be corrected by the application of operators conditioned error syndromes: 
\includegraphics{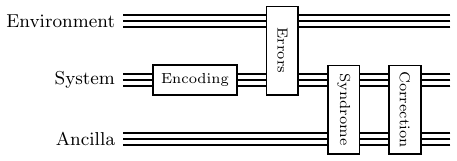}

Errors generally introduce entanglement between the system and the environment. After correction, this entanglement should have been transferred from the system to the ancillae, so that the environment and ancillae are entangled while the system is separable. Generically, this means that branching may occur in the combined state of the environment and ancillae, but not in the system. Equivalently, the state of the system should be the same in all branches. 
\begin{equation} |\text{Env}\rangle \otimes |\text{Sys}\rangle \otimes |\text{Anc}\rangle\end{equation}
\begin{equation} \text{After errors }\rightarrow \sum_n|\text{Env}_n^\prime\rangle \otimes |\text{Sys}_n^\prime\rangle \otimes |\text{Anc}\rangle\end{equation}
\begin{equation} \text{After correction }\rightarrow \sum_n|\text{Env}_n^{\prime\prime}\rangle  \otimes |\text{Sys}\rangle \otimes |\text{Anc}_n^\prime\rangle \end{equation}
This creates a similar situation as we considered in example \ref{sec:example:separable} where we saw branching independent of a separable system, and example \ref{sec:example:entangled} where we saw that branching becomes increasingly favored as long-range entanglement grows (between the environment and the ancillae in this case).

\section{Irreversibility}\label{sec:irreversibility}
Consider the case in which an initial state $|\Psi_0\rangle$ evolves into a state with branches: 
\begin{gather}|\Psi_0\rangle \longrightarrow |\Psi_t\rangle,\\
|\Psi_t\rangle = \sum_\vi c_\vi |\psi_\vi\rangle,\\
\CI(|\psi_\vi\rangle,\, |\psi_\j\rangle,\, \epsilon) - \CD(|\psi_\vi\rangle,\, |\psi_\j\rangle,\, 1-\epsilon)\gg1\; \forall\; \vi\neq\j. \end{gather}
Suppose the initial state $|\Psi_0\rangle$ has low entanglement between distant regions. For example, it may be a product state in a spatially-local basis, or only a constant small-depth circuit (or short time evolution) away from such a product state.

In such cases, the interference complexity between branches approximately lower-bounds the complexity of reversing $|\Psi_t\rangle$ to  $|\Psi_0\rangle$:
\begin{equation}\label{eq:irreversibility}
    \CI \left(|\psi_\vi\rangle,\, |\psi_\j\rangle,\, \Delta \right) \lesssim \C\left( |\Psi_t\rangle,\, |\Psi_0\rangle,\, \Delta \right) \text{ for all } \Delta.
\end{equation}
In other words, infeasibility of interference experiments implies irreversibility, since reversing $|\Psi_t\rangle$ to the initial product state $|\Psi_0\rangle$ is at least as hard as getting relative-phase information. 

This result does not assume any of the ``second law of complexity'' conjectures discussed in section \ref{sec:time_evolution}. These conjectures directly imply a stronger and more universal form of irreversibility, akin to the second law of thermodynamics for the increase of $\C\left( |\Psi_t\rangle, |\Psi_0\rangle \right)$ with time~\cite{brown2018second}. \\

    \textbf{Proof of equation (\ref{eq:irreversibility}):}\\
        Interference and distinguishability complexities are related by a simple change of basis, by construction. Detecting interference between $|\psi_\vi \rangle$ and $|\psi_\j \rangle$ is equivalent to distinguishing between $\frac{|\psi_\vi \rangle+e^{i\theta}|\psi_\j \rangle}{\sqrt{2}}$ and $\frac{|\psi_\vi \rangle-e^{i\theta}|\psi_\j \rangle}{\sqrt{2}}$ at some $\theta$:
\begin{align}\label{eq:CI_CD_related}
\begin{split}
&\CI(|\psi_\vi \rangle,\ |\psi_\j \rangle,\  \Delta)\\
&= \CD\left(\frac{|\psi_\vi \rangle+e^{i\theta}|\psi_\j \rangle}{\sqrt{2}},\ \frac{|\psi_\vi \rangle-e^{i\theta}|\psi_\j \rangle}{\sqrt{2}},\  \Delta\right).
\end{split}
\end{align}
Additionally, distinguishability complexity between orthogonal states is bounded from above by the complexity of transforming either state into a product state. So assuming $|\Psi_0\rangle$ is a product state:
\begin{align}\label{eq:distinguishability_bounded}
\begin{split}
    &\CD(\frac{|\psi_\vi \rangle+e^{i\theta}|\psi_\j \rangle}{\sqrt{2}},\ \frac{|\psi_\vi \rangle-e^{i\theta}|\psi_\j \rangle}{\sqrt{2}},\  \Delta) \\
& \leq \min \left[\C {\left(\frac{|\psi_\vi \rangle+e^{i\theta}|\psi_\j \rangle}{\sqrt{2}}, |\Psi_0\rangle, \Delta \right)}\right. ,\\ 
& \left. \hspace{1.3cm}  \C {\left(\frac{|\psi_\vi \rangle-e^{i\theta}|\psi_\j \rangle}{\sqrt{2}},  |\Psi_0\rangle, \Delta \right)}\right].
\end{split}
\end{align}

Finally, since $|\Psi_t\rangle = \sum_\vi c_\vi |\psi_\vi\rangle$, the relative complexity $\C\left( |\Psi_t\rangle,\  |\Psi_0\rangle,\  \Delta \right)$ must be at least as large as the right hand side of this inequality, and therefore at least as large as $\CI \left(|\psi_\vi\rangle,\  |\psi_\j\rangle,\  \Delta \right)$.

If  $|\Psi_0\rangle$ is not exactly a product state, but takes some number of gates $\C(|\Psi_0\rangle)$ to prepare from a product state, then $\C(|\Psi_0\rangle)$ is added to the right-hand side of equation (\ref{eq:distinguishability_bounded}), and the final inequality becomes 
\begin{equation}
    \CI \left(|\psi_\vi\rangle,\, |\psi_\j\rangle \right) \leq \C\left( |\Psi_t\rangle,\, |\Psi_0\rangle \right) + \C(|\Psi_0\rangle).
\end{equation}
However this extra term is relatively small and may be neglected in cases where the interference complexity is very high compared to the initial complexity ($\CI \left(|\psi_\vi\rangle,\; |\psi_\j\rangle \right) \gg \C(|\Psi_0\rangle)$), as will be the case if $|\Psi_0\rangle$ is only a constant small-depth circuit away from a product state while $|\psi_\vi\rangle$ and $|\psi_\j\rangle$ satisfy the definition of good branches (definition \ref{def:valid_branch}).

\section{Redundancy and robustness to coarse-graining}\label{sec:robustness}
One area where our definitions could be questioned is in their robustness to spatial coarse-grainings. Consider branching a ``single distinguishing qubit'' state: 
\begin{equation}\label{eq:single_distinguishing_qubit}\alpha|0\rangle|a\rangle + \beta|1\rangle|b\rangle\longrightarrow[\alpha|0\rangle|a\rangle,\ \beta|1\rangle|b\rangle]\end{equation}
where $|a\rangle$ and $|b\rangle$ are two states such that $\CI(|a\rangle,|b\rangle)\approx\CD(|a\rangle,|b\rangle)\gg1$ (for example they may be large Haar random states). The total interference complexity of the proposed branching is $\CI(|0\rangle|a\rangle,|1\rangle|b\rangle)\approx\CI(|a\rangle,|b\rangle)\gg1$,  but because of the distinguishing qubit the total distinguishability complexity is $\CD(|0\rangle|a\rangle,|1\rangle|b\rangle)=O(1)$. The condition for branching is satisfied, but only because of that single qubit. Note that this is the same example which we considered in equation \ref{eq:uniqueness_counterexample} of section \ref{sec:uniqueness} on the lack of a unique preferred basis.

This branching is obviously not robust to coarse-grainings, because that single qubit is likely to be lost. To ameliorate this, our definitions could be changed to require which-branch distinguishing information to be \textit{redundantly} accessible, rather than just accessible. This would bring our definition much more into line with quantum Darwinism style approaches which emphasize the importance of redundantly stored local records of ``which branch'' information~\cite{riedel2016objective, riedel2017classical}. Without such redundant records of events, how are different agents to agree on a shared classical reality, after all?

However it is not clear to us that a requirement for redundancy is entirely necessary for our purposes. Consider again the connection to error-correcting codes: a single distinguishing qubit entangled with logical code states like this is very likely to completely disrupt the code, so long as that single qubit is allowed to interact at all with the environment or undergo any undesired time evolution. Unless this physical degree of freedom is perfectly isolated and frozen, all of our arguments from the previous sections still hold, and distinguishing $\rho$ from $\rho_{\text{diag}}$ will be infeasible. A single qubit really can have a big effect.

Furthermore, we may find that the costs of enforcing redundancy are too high to pay. The cost is that the condition for branching is stricter, and so branches are defined less frequently, and can absorb less entanglement. If branches do not absorb entanglement fast enough in the relevant systems then they also cannot account for the apparently classical nature of the macroscopic world. Indeed, Riedel has recently argued on these grounds against definitions of branches which require redundant records~\cite{riedel2024generalizing}.

\section{Miscellaneous properties of $\CI$ and~$\CD$}\label{sec:misc_properties}
It is relatively easy to verify the following properties of $\CI$, $\CD$, and $\CR$ (assuming we are considering complexities between orthogonal states):

All of the complexities we have defined (and their proxies) are monotonically increasing functions of $\Delta$:
\begin{gather}
    \C_\textrm{[\normalfont{I,D,R}]}(\ket{a},\ket{b}, \Delta) \leq \C_\textrm{[\normalfont{I,D,R}]}(\ket{a},\ket{b},\,  \Delta+|\delta|\,),
\end{gather}
likewise, they are all symmetric:
\begin{gather}
    \C_\textrm{[\normalfont{I,D,R}]}(\ket{a},\ket{b}, \Delta) = \C_\textrm{[\normalfont{I,D,R}]}(\ket{b},\ket{a}, \Delta),
\end{gather}
and they are all unaffected by relative-phase: 
\begin{gather}
    \C_\textrm{[\normalfont{I,D,R}]}(\ket{a},\ket{b}, \Delta) = \C_\textrm{[\normalfont{I,D,R}]}(\ket{a},e^{i\theta}\ket{b}, \Delta).
\end{gather}

Distinguishability complexity is greater than or equal to interference-complexity in the conjugate basis: 
\begin{equation} \CD(\ket{a},\ket{b}) \geq  \CI\left(\frac{\ket{a}+\ket{b}}{2},\frac{\ket{a}-\ket{b}}{2}\right)  \end{equation}

The interference complexity proxy is bounded above and below by the relative state complexity:
\begin{equation}\label{eq:CI_bound}
\CR(|a \rangle,|b \rangle, \Delta/2) 
\leq 
\CIprox(|a \rangle,|b \rangle, \Delta/2) \leq \CR(|a \rangle,|b \rangle, \Delta),
\end{equation}
so the relative-state complexity may serve as an even simper proxy for $\CI$, especially when $\Delta$ is small. For the proof from below, swapping two states is strictly harder than mapping one state to another. For the proof from above, $\DeltaI/2 = \left|\bra{a}U\ket{b}\right| + \left|\bra{b}U\ket{a}\right|\geq\left|\bra{a}U\ket{b}\right| = \DeltaR$. 

Distinguishability complexity is bounded from above by the relative complexity of either state with a product state such as $|0\cdots 0\rangle$:
\begin{equation}\label{eq:CD_bound}
\CD(|a \rangle,|b \rangle) 
\leq \min \left[ \CR\left(|0\cdots 0\rangle, |a\rangle \right) ,  \CR\left( |0\cdots 0\rangle, |b \rangle \right)\right].
\end{equation}

So a sufficient (but not necessary) condition for branching is $\CR(|a \rangle,|b \rangle, \epsilon) - \CR(|0\cdots 0\rangle, |a\rangle,  1-\epsilon) \gg 1$: if it takes many more gates to get $\epsilon$ overlap between branches than it takes to prepare one of the branches from a product state.

The relative-state complexity follows a sort of triangle inequality: being able to map from $\ket{a}$ to $\ket{b}$ and from $\ket{b}$ to $\ket{c}$ implies the ability to map from $\ket{a}$ to $\ket{c}$ (at a reduced accuracy). In the worst case where the terms maximally cancel, the accuracy is reduced from $\Delta$ to $2\Delta^2-1$, while in the average case, if the phases are random, the accuracy is only reduced from $\Delta$ to $\Delta^2$:
\begin{equation}
\begin{split}
    &\CR\left(\ket{a},\ket{c}, 
{\Delta}^2 \pm |1-\Delta^2|\right) \\
&\leq     \CR(\ket{a},\ket{b}, \Delta) \,+\, \CR(\ket{b},\ket{c}, \Delta).    
\end{split}
\end{equation}

\subsection{Properties involving more than two branches}\label{sec:properties_multiple_branches}
Being able to distinguish $\ket{a}$ from $\frac{\ket{b} + \ket{c}}{\sqrt{2}}$ at a high accuracy implies the ability to distinguish $\ket{a}$ from $\ket{b}$ with a lower accuracy, if $\ket{a}$, $\ket{b}$ and $\ket{c}$ are orthogonal:
\begin{equation}
    \CDprox\left(\ket{a}, \ket{b}, 1-2\epsilon \right) \leq \CDprox\left(\ket{a}, \frac{\ket{b} + \ket{c}}{\sqrt{2}}, 1-\epsilon \right),\label{eq:CD_three_simple}
\end{equation}
or more generally,
\begin{equation}
\begin{split}
&\CDprox\left(\ket{a}, \ket{b}, 1-\frac{\epsilon}{p} \right) \\
&\leq \CDprox\left(\ket{a}, \sqrt{p}\ket{b} + \sqrt{1-p}\ket{c}, 1-\epsilon \right),\label{eq:CD_three}  
\end{split}
\end{equation}
and likewise for distinguishing $\ket{a}$ from $\ket{c}$.

Being unable to even $\epsilon$-approximately interfere $\ket{a}$ with $\frac{\ket{b} + e^{i\theta}\ket{c}}{\sqrt{2}}$ at any $\theta$ implies being unable to interfere $\ket{a}$ with $\ket{b}$. Specifically,
\begin{equation}
\begin{split}
    \CIprox\left(\ket{a}, \ket{b}, \sqrt{2}\epsilon \right)
    &\geq \min_\theta\left[ \CIprox\left(\ket{a}, \frac{\ket{b} + e^{i\theta}\ket{c}}{\sqrt{2}}, \epsilon \right)\right],\label{eq:CI_three_simple}
\end{split}
\end{equation}
or more generally,
\begin{equation}
\begin{split}
    &\CIprox\left(\ket{a}, \ket{b}, \frac{\epsilon}{\sqrt{p}} \right)\\
    &\geq \min_\theta\left[ \CIprox\left(\ket{a}, \sqrt{p}\ket{b} + e^{i\theta}\sqrt{1-p}\ket{c}, \epsilon \right)\right],\label{eq:CI_three}
\end{split}
\end{equation}
and likewise for interfering $\ket{a}$ with $\ket{c}$.

\subsection{Compatibility of two decompositions of a three-branch state} \label{sec:three_branch_compatibility}
We can use the results of \ref{sec:properties_multiple_branches} to show a compatibility between different two-branch splittings and a three-branch splitting.
Let $\ket{\Psi} = \frac{1}{\sqrt{3}}\ket{a} + \frac{1}{\sqrt{3}}e^{i\theta}\ket{b} + \frac{1}{\sqrt{3}}e^{i\phi}\ket{c}$. Suppose that $[\frac{\ket{a}}{\sqrt{3}}, \frac{\ket{b} + e^{i\phi}\ket{c}}{\sqrt{3}}]$ is a good branch decomposition, and so is $[\frac{\ket{a} + e^{i\theta}\ket{b}}{\sqrt{3}},  \frac{\ket{c}}{\sqrt{3}}]$ for any phases $\theta$ and $\phi$. By equations (\ref{eq:CD_three_simple}) and (\ref{eq:CI_three_simple}) the three-branch splitting $[\frac{\ket{a}}{\sqrt{3}}, \frac{\ket{b}}{\sqrt{3}}, \frac{\ket{c}}{\sqrt{3}}]$ must also be a good branch decomposition. 

Specifically, if the ``branchiness'' of the first splitting is $B_1\coloneqq$
\begin{equation}  
\min_\phi\CI\left(\ket{a}, \frac{\ket{b}+e^{i\phi}\ket{c}}{\sqrt{2}}, \epsilon \right) - \CD\left(\ket{a}, \frac{\ket{b}+\ket{c}}{\sqrt{2}}, 1-\epsilon \right) \end{equation} 
and the ``branchiness'' of the second splitting is $B_2\coloneqq$
\begin{equation}  \min_\theta \CI\left(\frac{\ket{a}+e^{i\theta}\ket{b}}{\sqrt{2}}, \ket{c}, \epsilon \right) - \CD\left(\frac{\ket{a}+\ket{b}}{\sqrt{2}}, \ket{c}, 1-\epsilon \right)
\end{equation} 
then it must be true that
\begin{gather}  \CI\left(\ket{a}, \ket{b}, \sqrt{2}\epsilon \right) - \CD\left(\ket{a}, \ket{b}, 1 -2\epsilon \right) \geq B_1,\\
 \CI\left(\ket{b}, \ket{c}, \sqrt{2}\epsilon \right) - \CD\left(\ket{b}, \ket{c}, 1-2\epsilon \right) \geq B_2,\end{gather} 
and
\begin{equation}  \CI\left(\ket{c}, \ket{a}, \sqrt{2}\epsilon \right) - \CD\left(\ket{c}, \ket{a}, 1-2\epsilon \right) \geq \max(B_1, B_2), \end{equation} 
so if both of the bipartite branchings are good ($B_1\gg 1$ and $B_2\gg1$) for all $\theta$ and $\phi$, then the combined tripartite branching must also be good. Similar results hold if the branches have different relative weights, using equations (\ref{eq:CD_three}) and (\ref{eq:CI_three}).

\end{document}